\documentclass{article}
\usepackage[outer=25mm,inner=25mm,vmargin=10mm,includehead,includefoot,headheight=15pt]{geometry}
\usepackage{lipsum}
\usepackage{multicol}
\usepackage{graphicx}
\usepackage{caption}
\usepackage{hyperref} 
\usepackage[english]{babel}

\usepackage{colortbl}
\usepackage{color}
\definecolor{pink}{rgb}{1.,0.75,0.8}
\definecolor{green}{rgb}{0.3,1,0.3}
\definecolor{dgreen}{rgb}{0.,0.6,0.}
\definecolor{gold}{rgb}{1.,0.84,0.}
\definecolor{beige}{rgb}{0.96,0.96,0.86}
\definecolor{myyellow}{rgb}{1.,0.84,0.8}
\definecolor{grey}{rgb}{0.8,0.8,0.8}
\definecolor{darkyellow}{cmyk}{0,0,0.5,0.5}
\definecolor{darkwhite}{gray}{0.1}
\definecolor{lightblack}{gray}{0.9}
\definecolor{lightblue}{rgb}{0.1,0.4,1.0}
\definecolor{fucsia}{rgb}{1.,0.4,0.9}

\newenvironment{Figure}
  {\par\medskip\noindent\minipage{\linewidth}}
  {\endminipage\par\medskip}

\begin{document}
\title{Study of Tritium Production and Interactions in LiD}

		\author{Serena Fattori\footnote{Email: \href{mailto:dr.serena.fattori@gmail.com}{dr.serena.fattori@gmail.com}, ORCID: \href{https://orcid.org/0000-0002-9381-7620}{0000-0002-9381-7620}} , Rino Persiani\thanks{Email: \href{mailto:rinopersiani@gmail.com}{rinopersiani@gmail.com}, ORCID: \href{https://orcid.org/0000-0002-3100-1466}{0000-0002-3100-1466}}, Ugo Abundo}
\date{31st March 2017}
\maketitle

\begin{abstract}
This study focuses on a fuel cell composed of Lithium Deuteride (LiD) in a spherical geometry, in which isotropic monoenergetic neutrons of 0.025~eV (thermal neutrons) are generated at the center. The objective is to investigate the production of Tritium via interactions with Lithium-6. The physics of the process has been modeled and analyzed for thirteen different concentrations of Lithium-6 to determine the most advantageous option among them. Monte Carlo simulations were performed using the GEANT4 toolkit to analyze neutron interactions and tritium production efficiency as a function of Lithium-6 concentration.
\end{abstract}

\begin{multicols}{2}
\section{Introduction} \label{sec:INT}
The purpose of this second phase of the work (for the first part refer to \cite{Report1}) is to study a fuel cell composed of Lithium Deuteride in a spherical geometry, at the center of which isotropic monoenergetic neutrons of 0.025~eV (thermal neutrons) are generated. The objective is to verify the production of Tritium through reactions with Lithium-6. The physics of the process has been implemented and studied for thirteen different concentrations of Lithium-6 to identify the most efficient among all those considered for comparison.

\section{Materials and Methods} \label{sec:MM}
The study of neutron interaction processes in Lithium Deuteride, along with the analysis of the fuel cell model, was carried out through Monte Carlo simulations using the GEANT4 toolkit \cite{GEANT4}, version 10.2.

\subsection{Geometry}
The studied geometry for the fuel cell consists of a sphere with a point-like neutron source at its center. The sphere's radius is set at 20.0~cm to ensure that all primary neutrons are captured, regardless of the Lithium-6 concentration considered.

\begin{figure*}[t]
\centering
	\begin{minipage}[t]{0.3\linewidth} 
		\centering
		 \includegraphics[width= 0.9\textwidth]{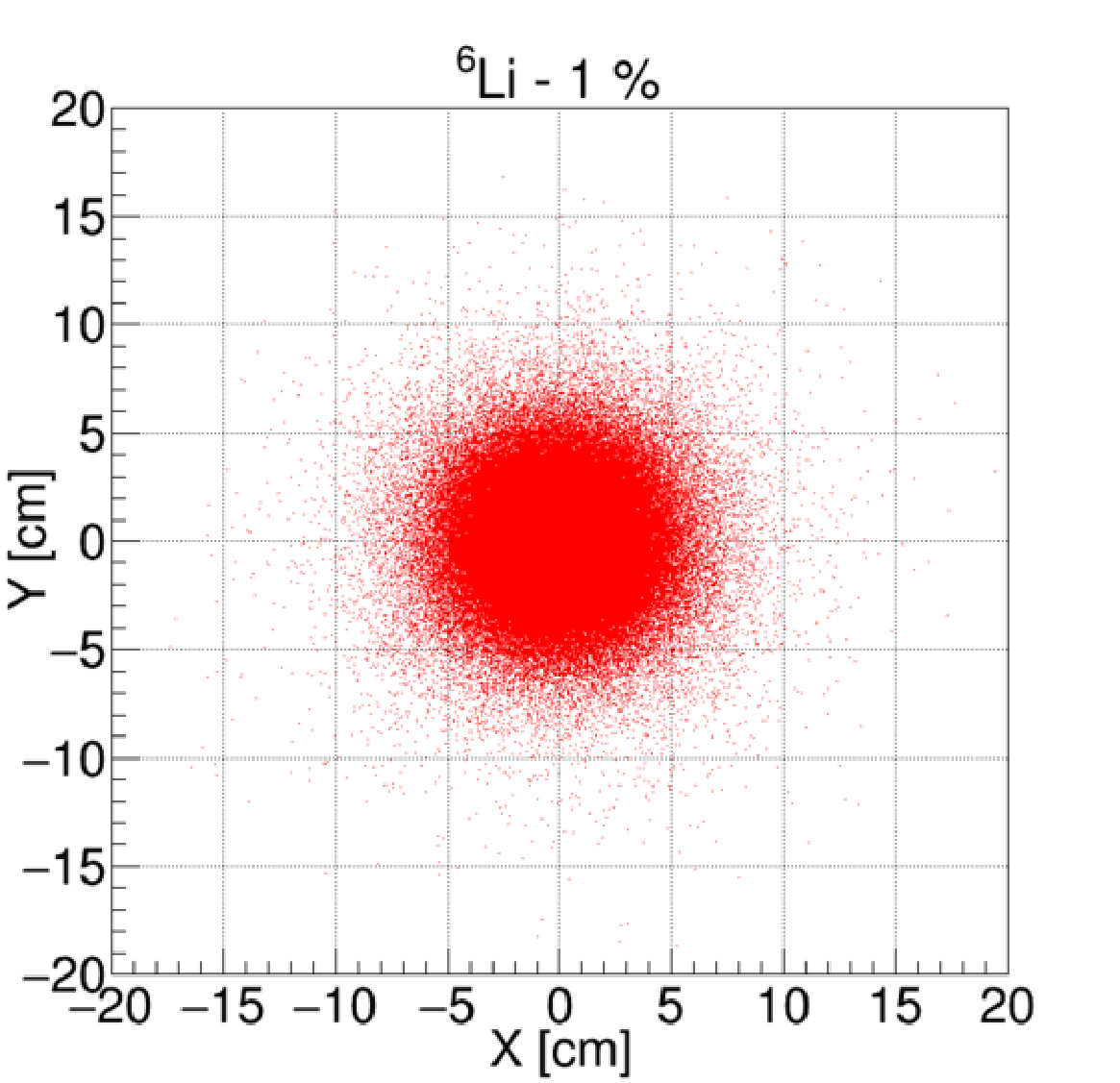} 
	\end{minipage}
		\hspace{0.5cm} 
	\begin{minipage}[t]{0.3\linewidth}
		\includegraphics[width= 0.9\textwidth]{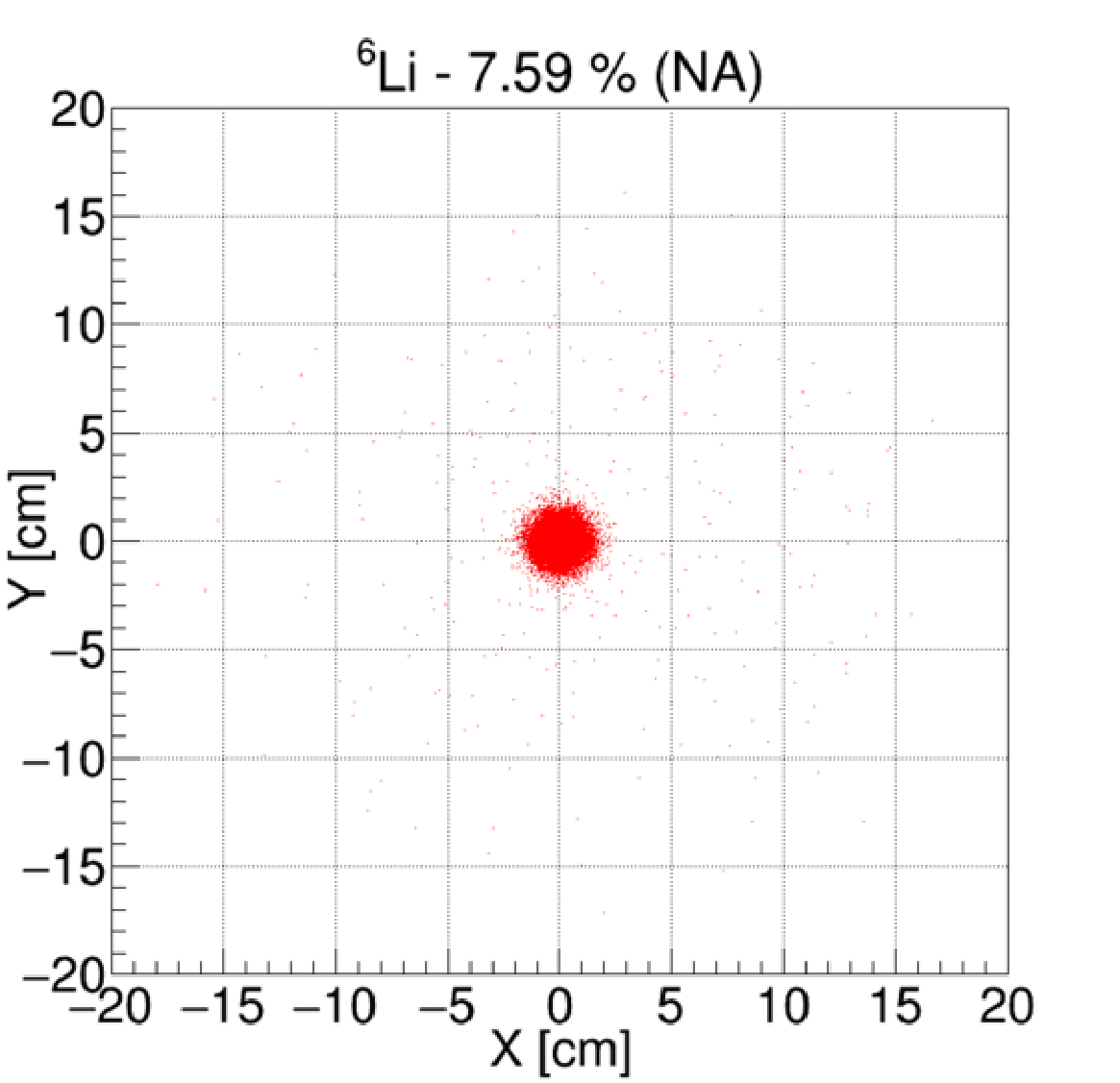} 	 
	\end{minipage}
	\begin{minipage}[t]{0.3\linewidth}
		\includegraphics[width= 0.9\textwidth]{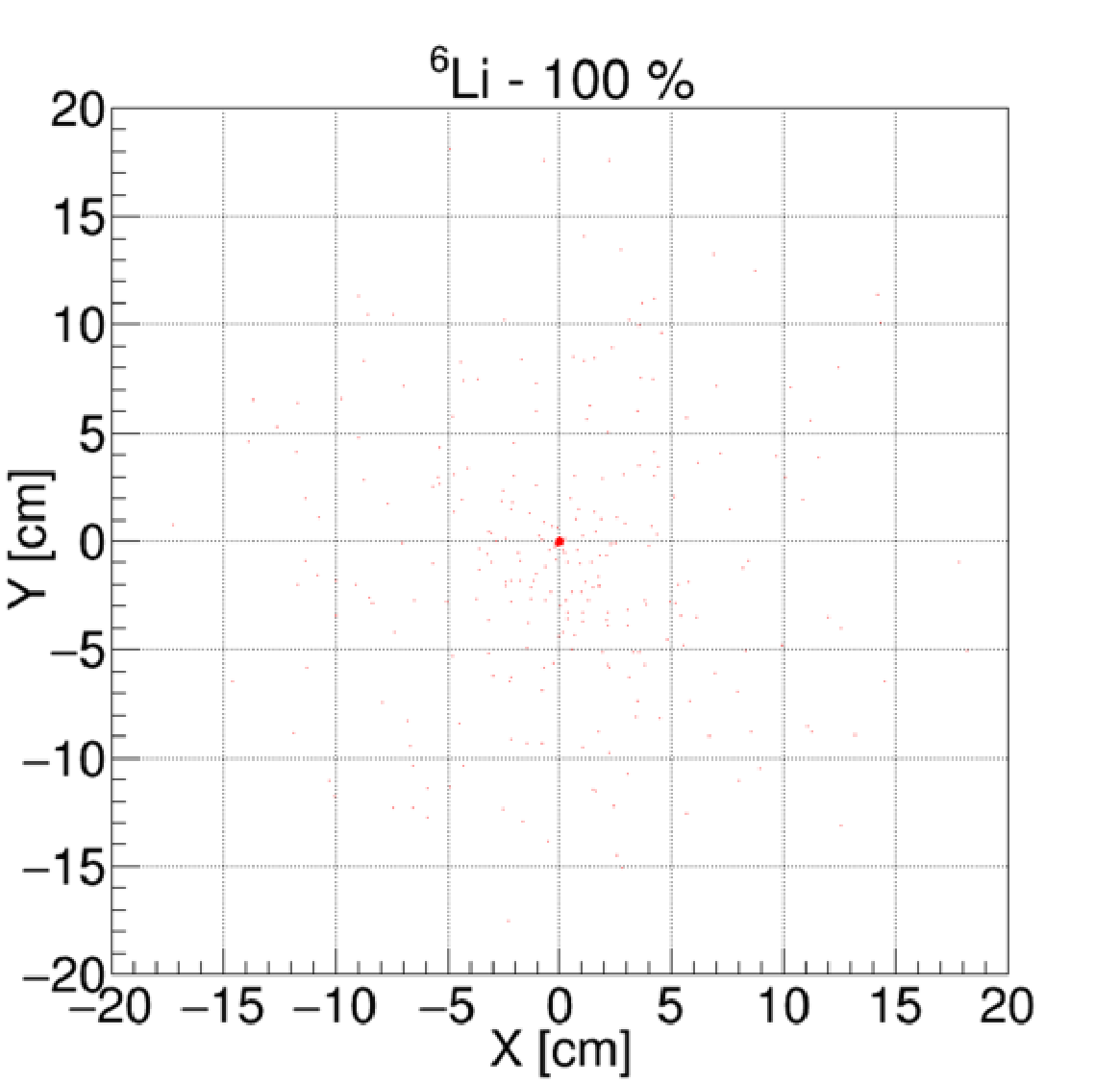} 	 
	\end{minipage}
	  \caption{Spatial distributions of primary neutron interaction points in three limiting cases: Lithium-6 concentration at 1\% (left image), Lithium-6 concentration at 7.59\% (center image), and Lithium-6 concentration at 100\% (right image).}
  \label{3Sfere}
\end{figure*}

Figure~\ref{3Sfere} illustrates the spatial distributions of the primary neutron interaction points for three cases:
\begin{itemize}
\item The first case (large sphere on the left) corresponds to a Lithium-6 concentration of 1\% (depleted compared to the natural concentration);
\item The second case (medium sphere in the center) represents a natural Lithium-6 concentration of 7.59\%;
\item The third case (small sphere on the right) corresponds to a Lithium-6 concentration of 100\% (maximum enrichment).
\end{itemize}
These images provide a preliminary qualitative insight: as the Lithium-6 concentration increases, a smaller thickness is required for all primary thermal neutrons to interact within the fuel cell. For quantitative details concerning these three cases and the remaining ten, refer to the subsequent sections of this report.

\subsection{Physics}
For the interaction modeling, the "High Precision" (HP) mode was selected, as in the previous study on neutron moderation \cite{Report1}. This choice provides the most realistic description of low-energy processes; however, it comes at the cost of a significant slowdown in interaction execution. For convenience, we recall the neutron physics models activated in this set of simulations, as previously described.

The primary neutron interaction processes considered include:
\begin{itemize}
    \item Elastic Scattering;
    \item Inelastic Scattering;
    \item Capture;
    \item Fission.
\end{itemize}

Various models were used for different energy ranges, with the "High Precision" mode employed for neutron transport and interactions below 20~MeV, based on experimental data-driven models. The following models were also used \cite{GEANT4models}:
\begin{itemize}
    \item hElasticCHIPS: describes hadron-nucleus elastic scattering using M.~Kossov’s parameterized cross-sections;
    \item QGSP: "Quark Gluon String model," a theoretical model for high-energy interactions;
    \item FTFP: models string formation in hadron-nucleus collisions;
    \item Binary Cascade: generates final states in inelastic hadron scattering;
    \item nRadCapture: describes neutron capture in the high-energy range;
    \item G4LFission: models high-energy fission processes.
\end{itemize}

\begin{Figure}
 \centering
  \includegraphics[width=\linewidth]{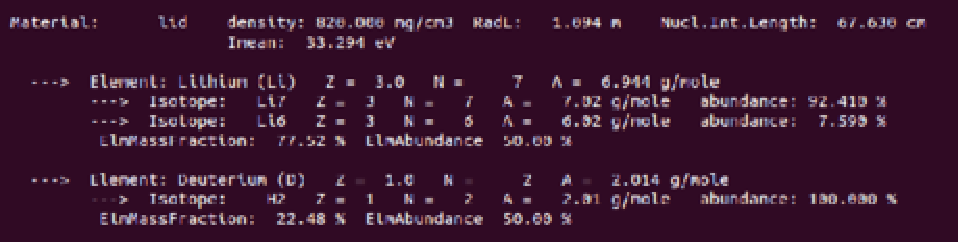}
 \captionof{figure}{Image captured from the Terminal during the execution of the simulation: material chemistry.}
 \label{Interazioni}
\end{Figure}

\section{Data and Analysis} \label{sec:DA}

\subsection{Neutron Interactions in the Fuel Cell}
To study Tritium production, $10^{6}$ neutrons of 0.025~eV were generated from the center of a sphere of 20.0~cm radius, composed of Lithium Deuteride. The primary reaction involved is:

\begin{equation}
    ^{6}Li + n \to  t + \alpha 
\end{equation}

The mean kinetic energy available to Tritium is 2.726~MeV, while for the alpha particle, it is 2.055~MeV (as shown in Figures~\ref{neuine_triton_energy} and~\ref{neuine_alpha_energy}).

The frequency of this inelastic interaction was then studied as a function of the sphere's radius for different Lithium-6 concentrations. The obtained results are illustrated in Figures~\ref{neuine_radius_007} (for the natural concentration), ~\ref{Spettri1} (for concentrations of 1\%, 5\%, 10\%, 20\%, 30\%, and 40\%), and~\ref{Spettri2} (for concentrations of 50\%, 60\%, 70\%, 80\%, 90\%, and 100\%). In the images, the contributions to the reaction from primary neutrons are shown in red, while the part of the distribution fed by secondary neutrons is shown in white.

From the analysis of the graphs, it is evident that:
\begin{itemize}
\item The contribution from primary neutrons follows an exponentially decreasing trend as a function of distance and drops rapidly (within 2~cm already for the first enrichment considered: 10\% Lithium-6);
\item The contribution from secondary neutrons remains constant as a function of distance.
\end{itemize}

The efficiency of this Tritium production process by thermal neutrons was verified as a function of Lithium-6 concentration, leading to the following results: efficiency $>99\%$ for a Lithium-6 concentration of 1\%, and efficiency $>99.9\%$ for all other tested concentrations ($\ge 5\%$).

The outgoing neutrons from the fuel sphere and their average energy were then analyzed as a function of Lithium-6 concentration. The results obtained are shown in Figure~\ref{exitneu_number} and Figure~\ref{exitneu_meanenergy}, respectively. 
The results clearly indicate that the number of neutrons leaving the sphere decreases as the Lithium-6 concentration increases, while their average energy increases accordingly.
\end{multicols}  
\newpage

 \begin{Figure}
	 \centering
	  \vspace{1.5cm}
 	\includegraphics[width=0.9\linewidth]{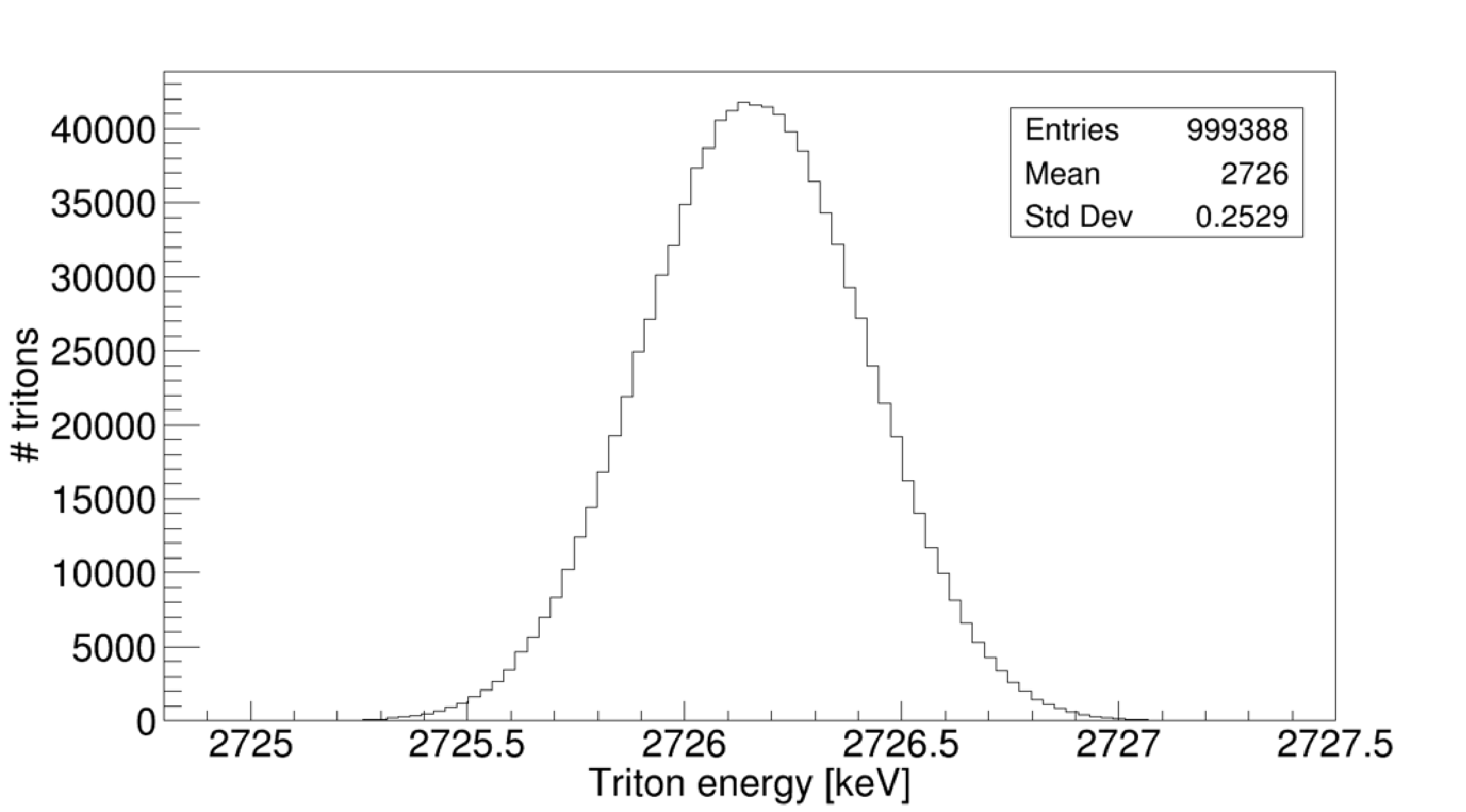}
 	\captionof{figure}{Energy spectrum available to Tritium, produced by the reaction $^{6}Li + n \to  t + \alpha$.}
 	\label{neuine_triton_energy}
 \end{Figure}
 
 \vspace{1.5cm}

\begin{Figure}
 	\centering
 	\includegraphics[width=0.9\linewidth]{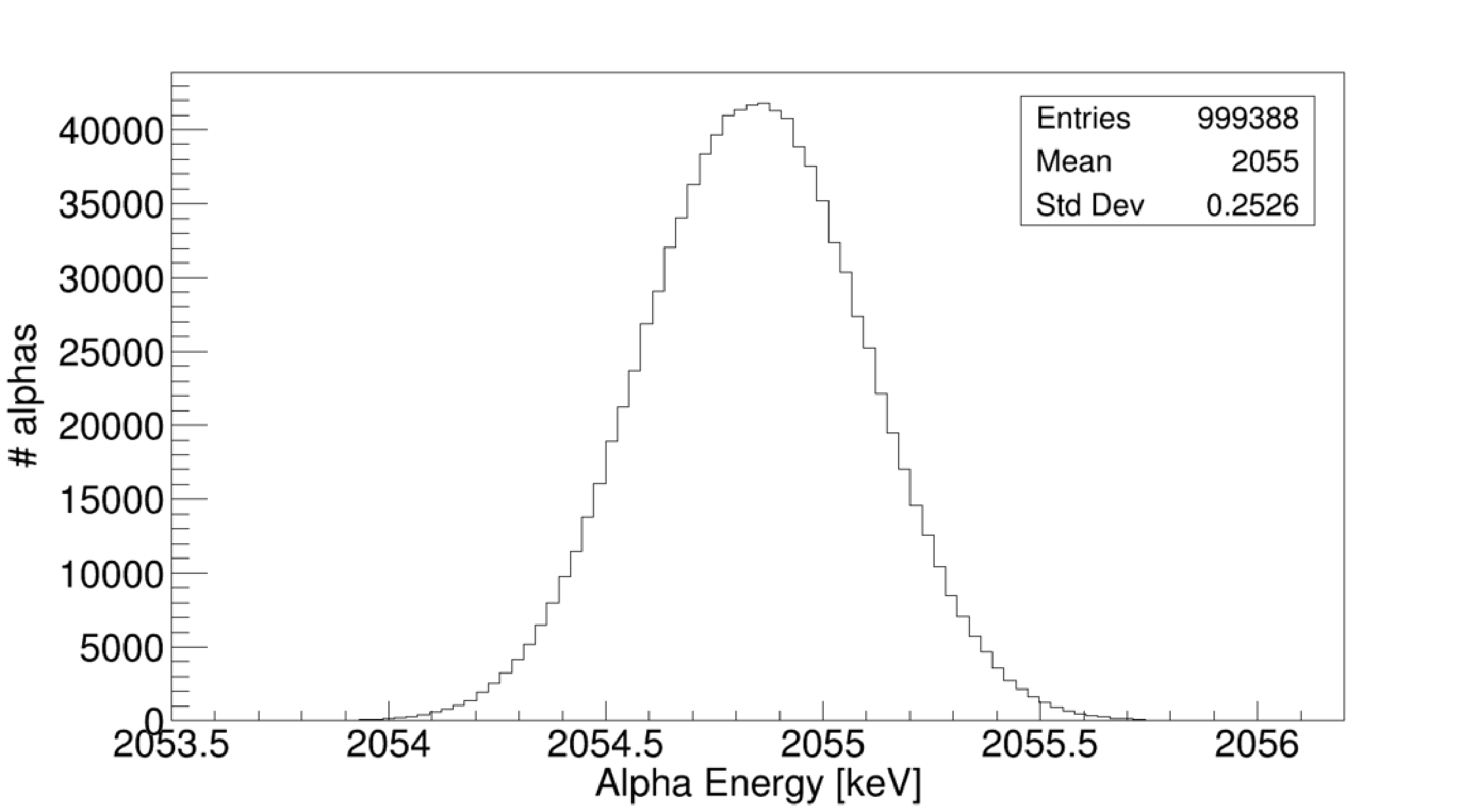}
 	\captionof{figure}{Energy spectrum available to $\alpha$ particle, produced by the reaction $^{6}Li + n \to  t + \alpha$.}
 	\label{neuine_alpha_energy}
 \end{Figure}

\begin{Figure}
 	\centering
	\vspace{3.0cm}
 	\includegraphics[width=0.9\linewidth]{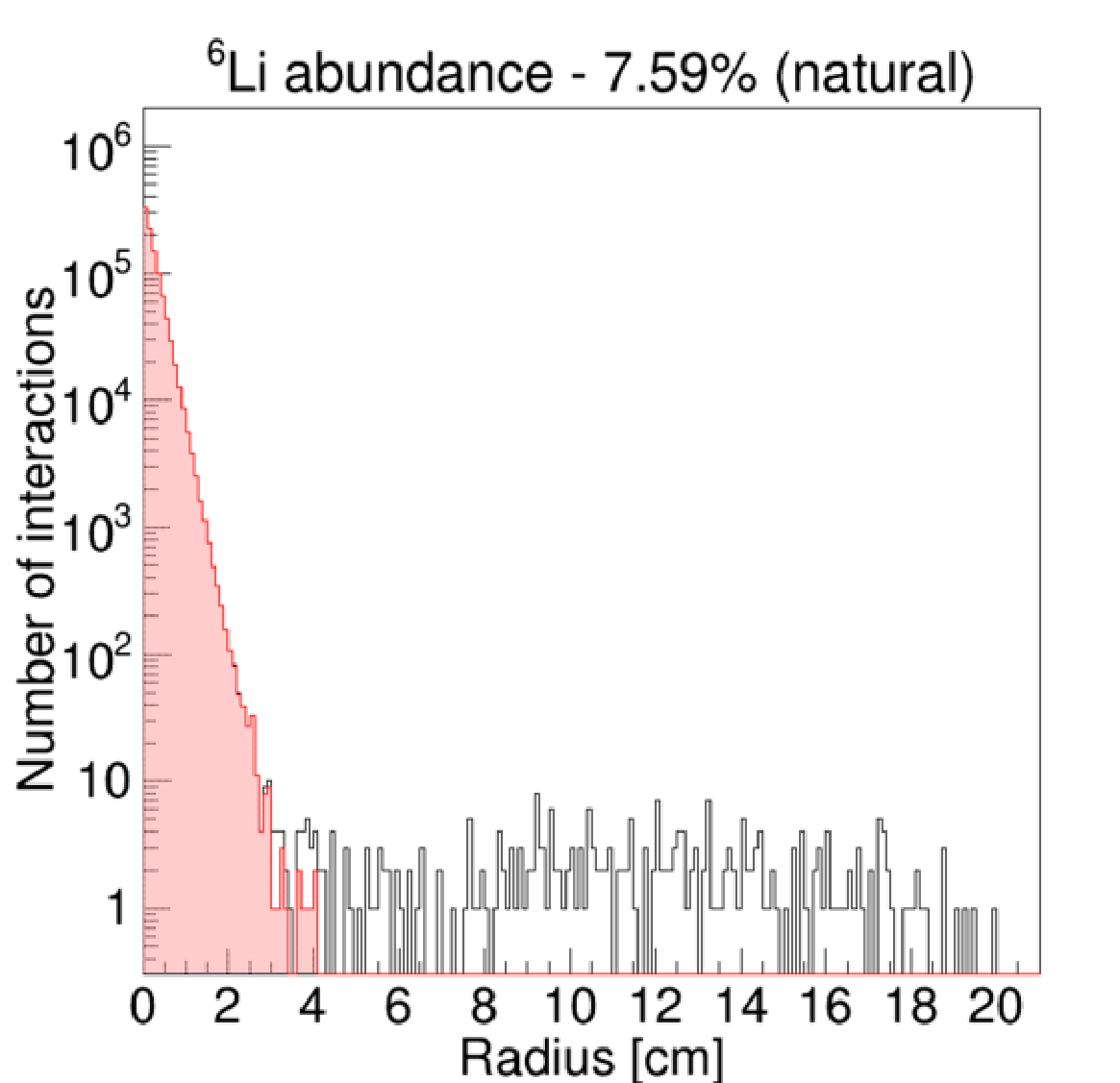}
 	\captionof{figure}{Frequency of the reaction $^{6}Li + n \to  t + \alpha$ as a function of the fuel sphere radius, for a natural concentration of Lithium-6.}
 	\label{neuine_radius_007}
\end{Figure}

\newpage
  
\begin{Figure}
\centering
\vspace{0.5cm}
	\begin{minipage}[t]{0.45\linewidth} 
		\centering
		 \includegraphics[width= 0.90\textwidth]{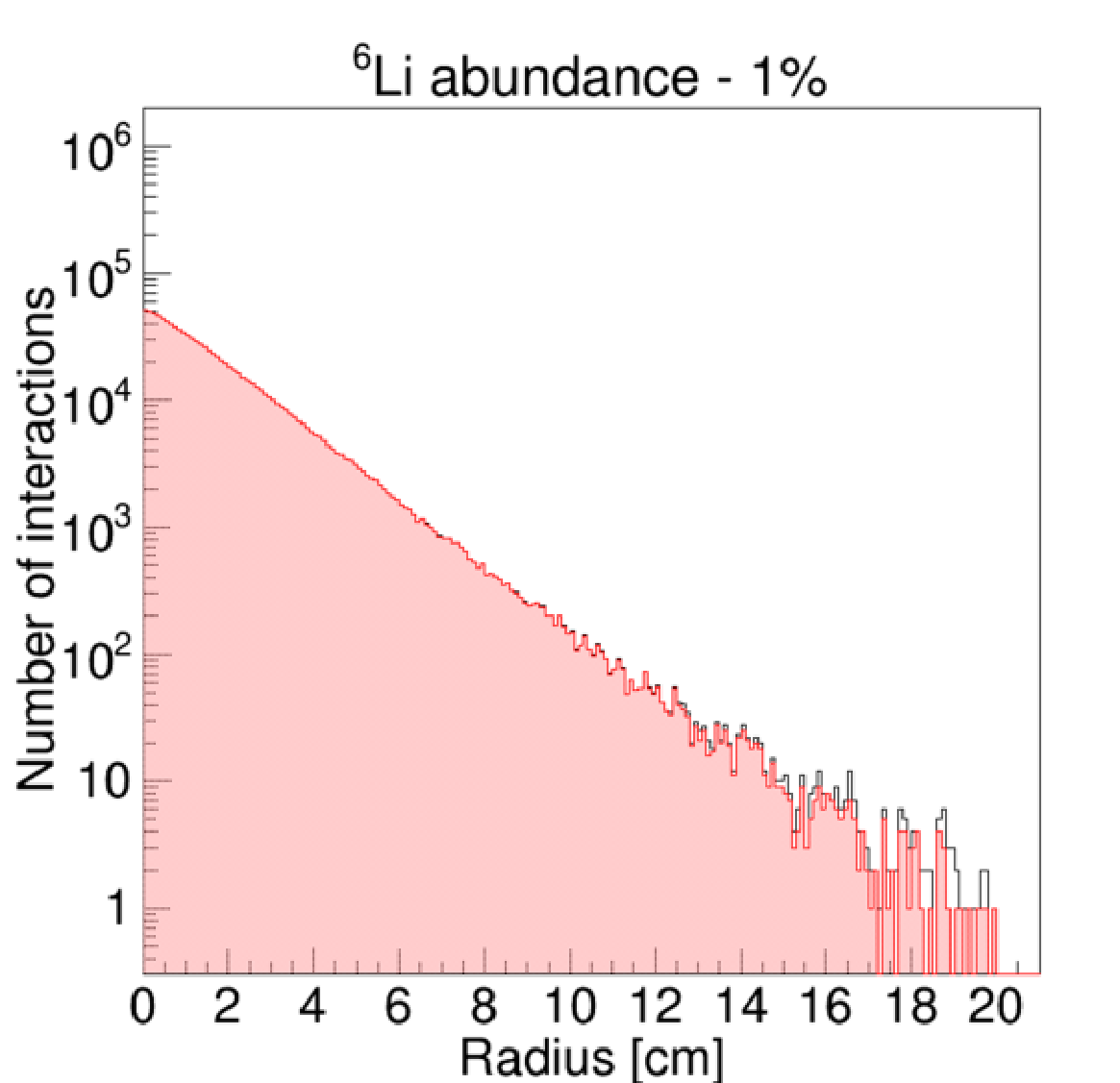} 
	\end{minipage}
		\hspace{0.5cm} 
	\begin{minipage}[t]{0.45\linewidth}
		\includegraphics[width= 0.90\textwidth]{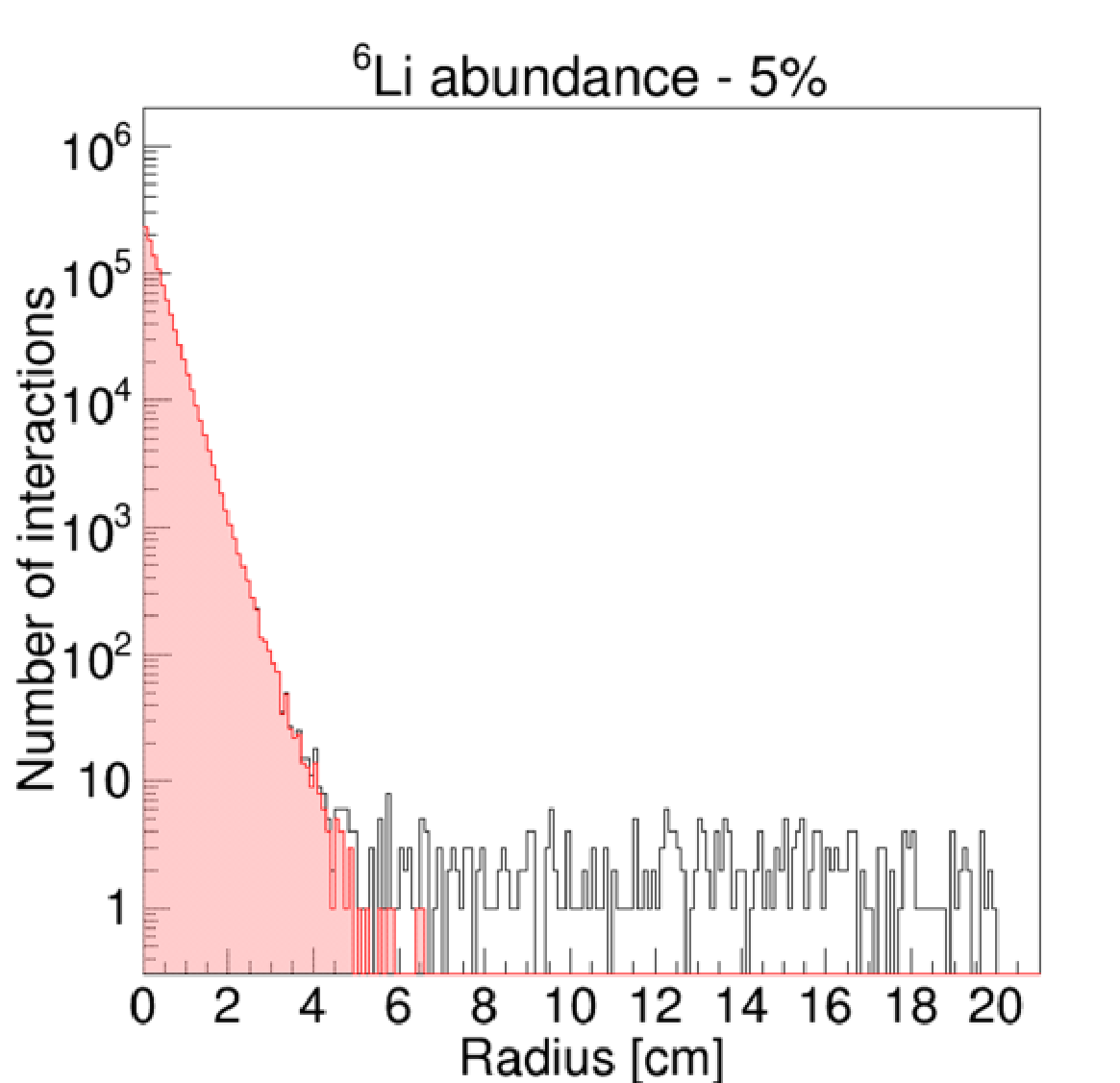} 
	\end{minipage}
 \end{Figure}

\begin{Figure}
\centering
	\begin{minipage}[t]{0.45\linewidth} 
		\centering
		 \includegraphics[width= 0.90\textwidth]{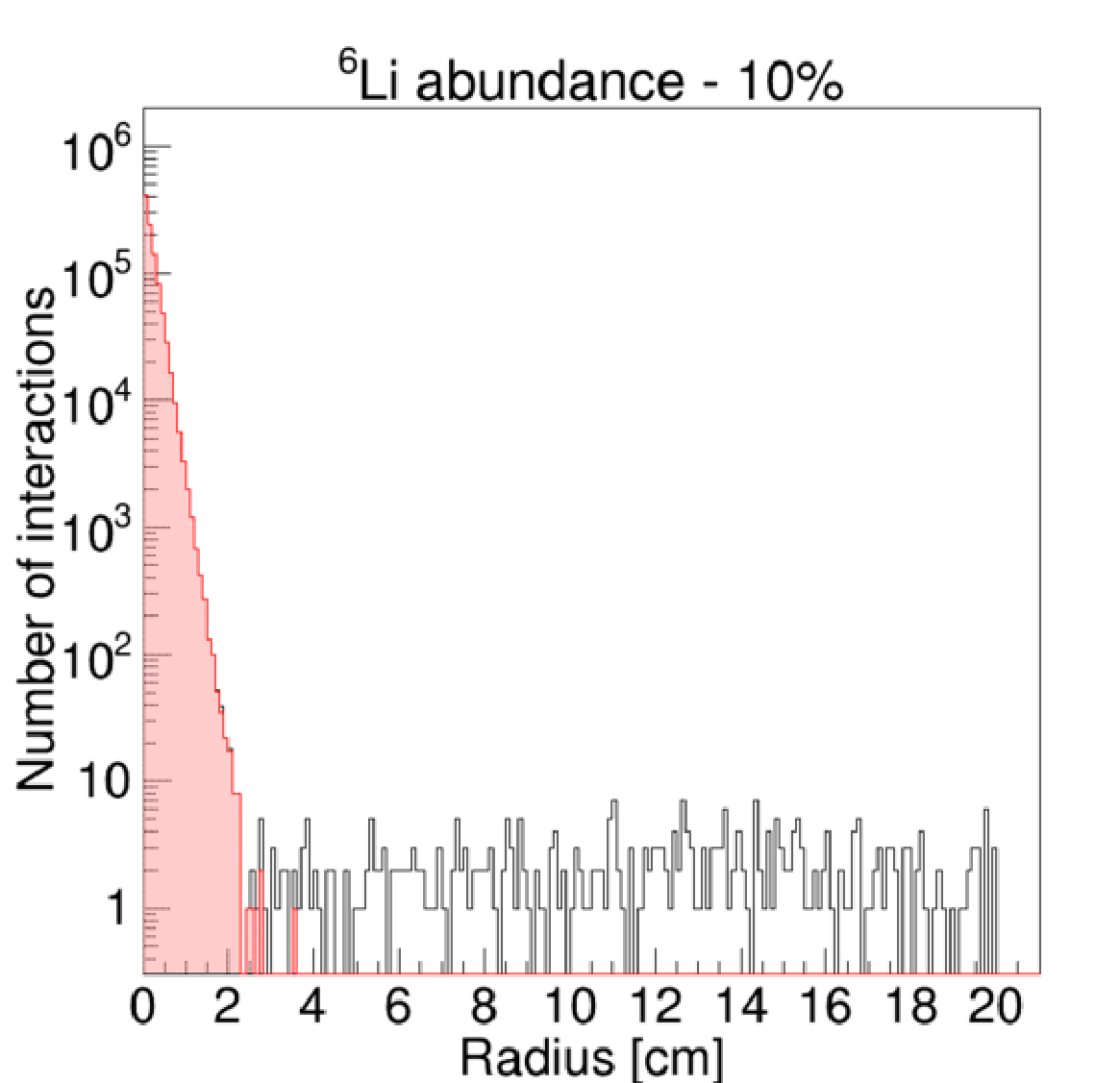} 
	\end{minipage}
		\hspace{0.5cm} 
	\begin{minipage}[t]{0.45\linewidth}
		\includegraphics[width= 0.90\textwidth]{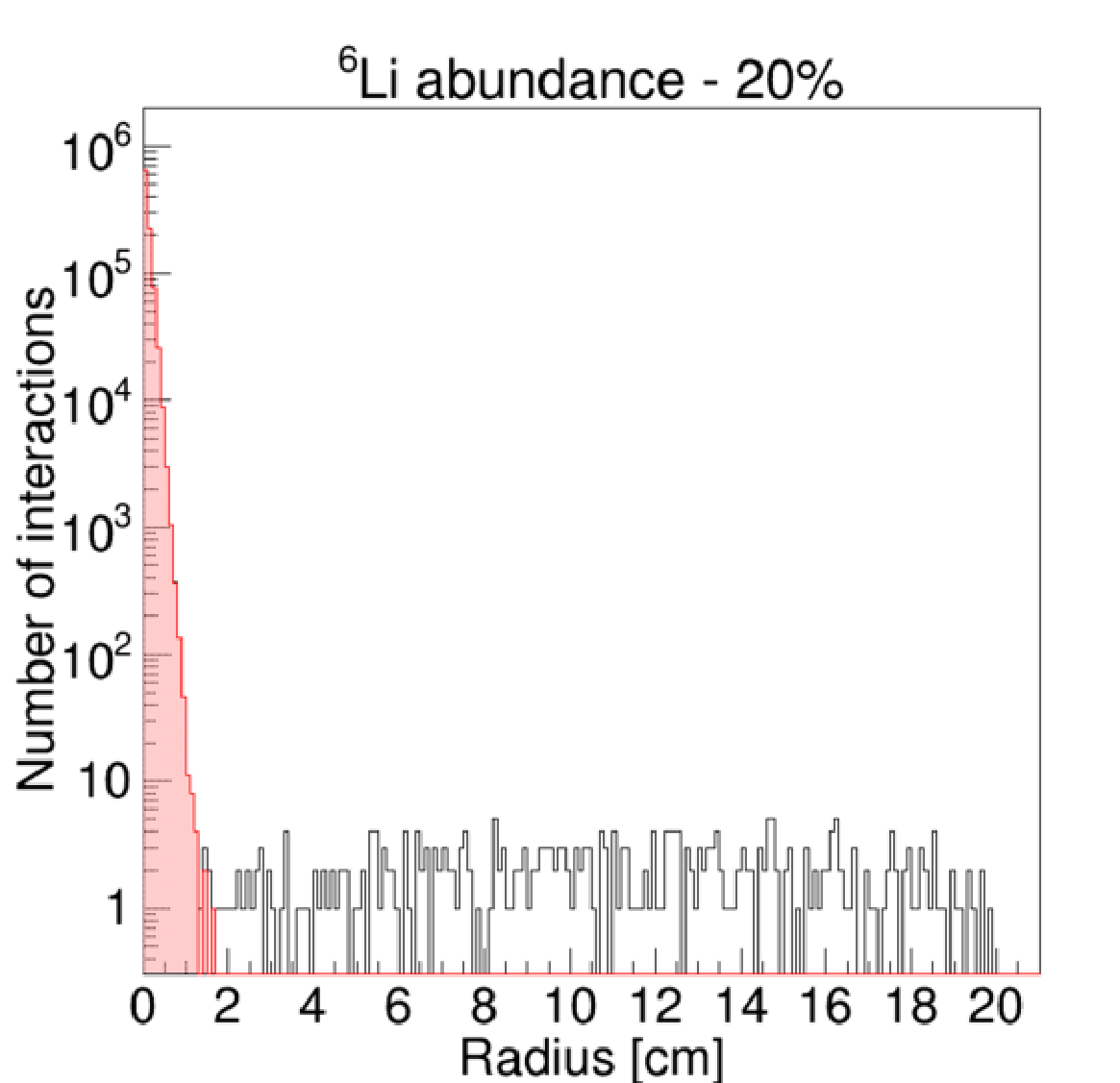} 		 
	\end{minipage}
 \end{Figure}

\begin{Figure}
\centering
	\begin{minipage}[t]{0.45\linewidth} 
		\centering
		 \includegraphics[width= 0.90\textwidth]{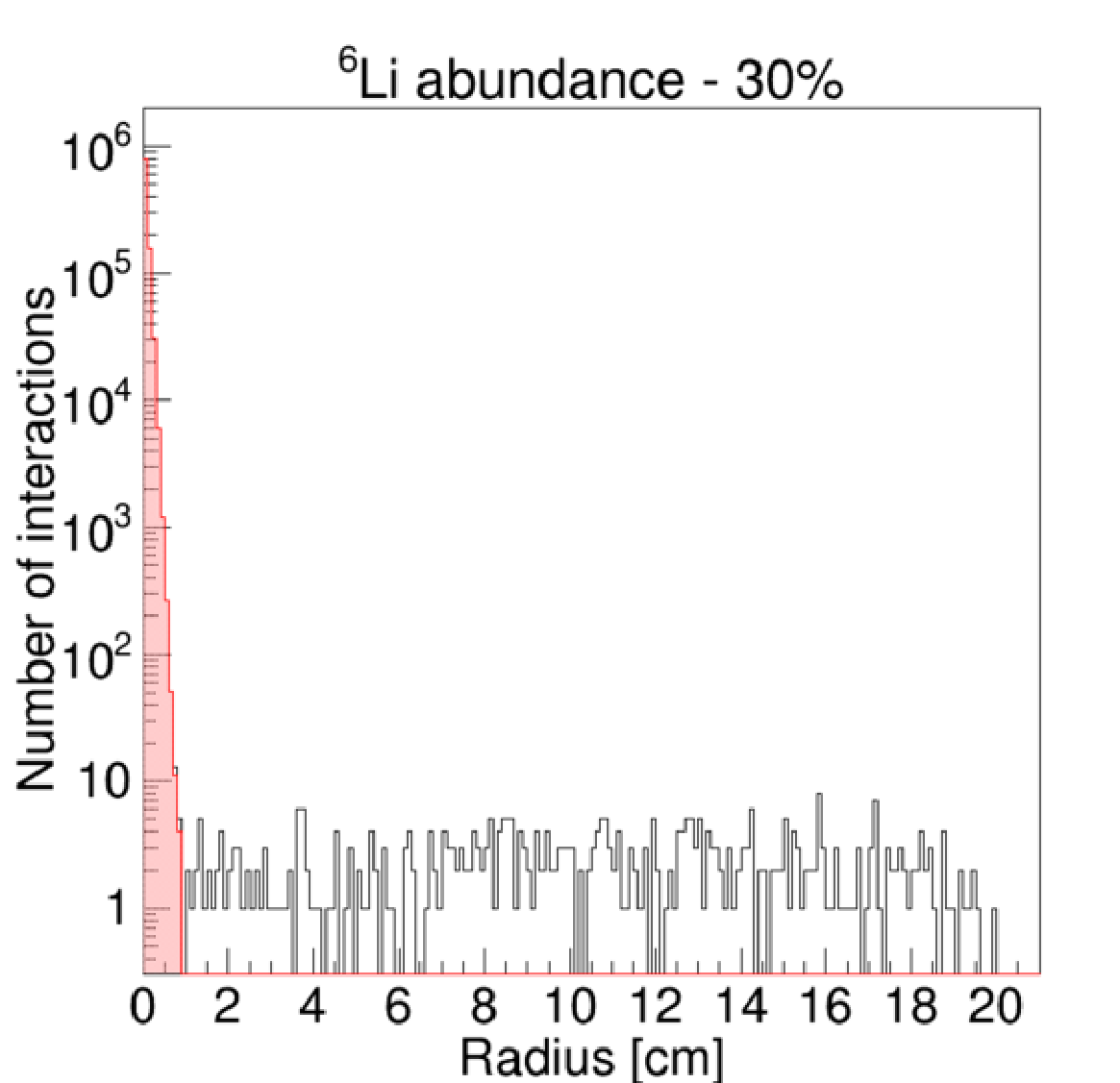} 
	\end{minipage}
		\hspace{0.5cm} 
	\begin{minipage}[t]{0.45\linewidth}
		\includegraphics[width= 0.90\textwidth]{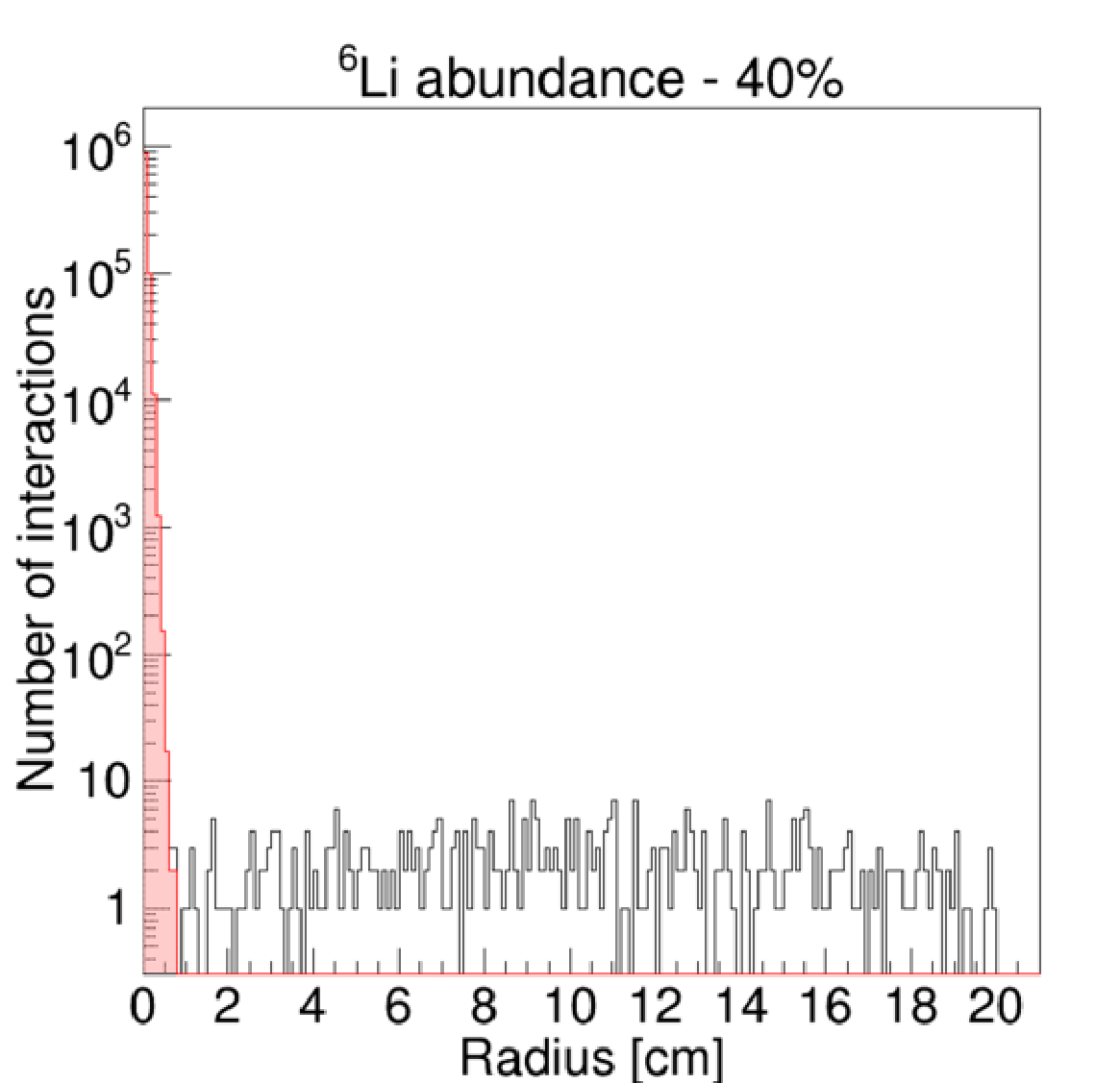} 	 
	\end{minipage}
	\captionof{figure}{Frequency of the reacion $^{6}Li + n \to  t + \alpha$ as a function of the fuel sphere radius, for Lithium-6 concentrations of 1\%, 5\%, 10\%, 20\%, 30\% e 40\%.}
	\label{Spettri1}
 \end{Figure}

\begin{Figure}
\centering
\vspace{0.5cm}
	\begin{minipage}[t]{0.45\linewidth} 
		\centering
		 \includegraphics[width= 0.90\textwidth]{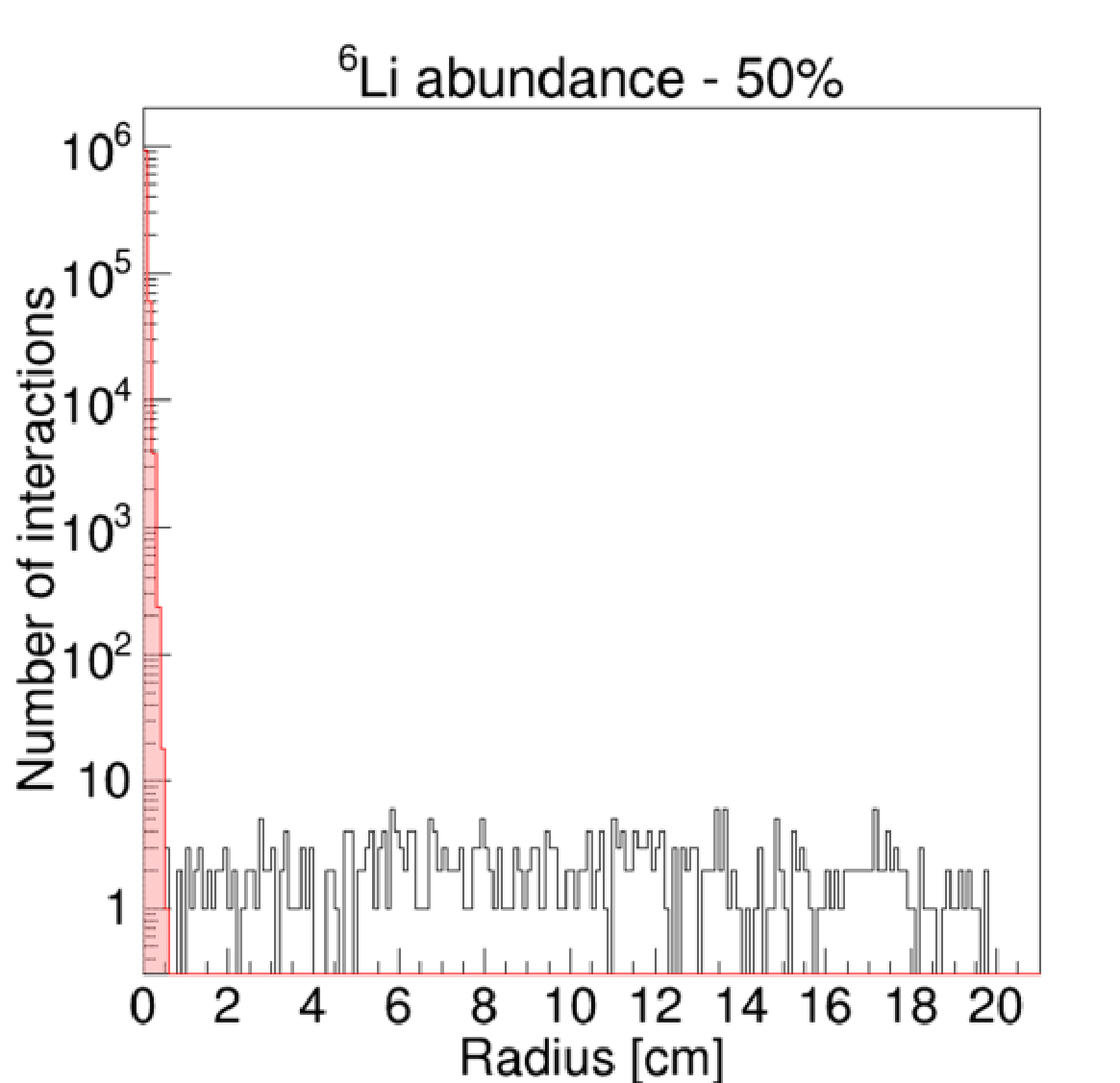} 
	\end{minipage}
		\hspace{0.5cm} 
	\begin{minipage}[t]{0.45\linewidth}
		\includegraphics[width= 0.9\textwidth]{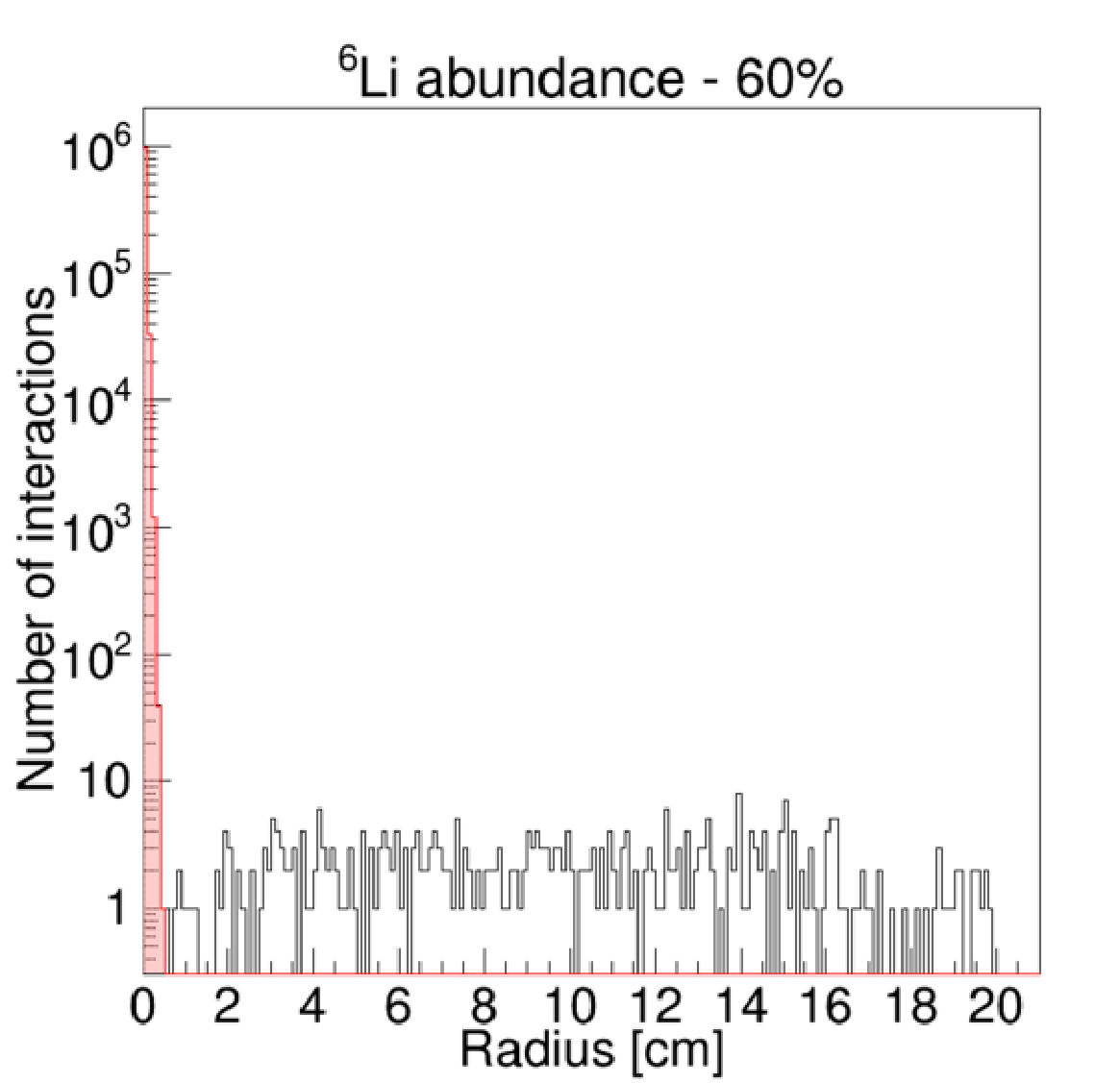} 	 
	\end{minipage}
 \end{Figure}

\begin{Figure}
\centering
	\begin{minipage}[t]{0.45\linewidth} 
		\centering
		 \includegraphics[width= 0.9\textwidth]{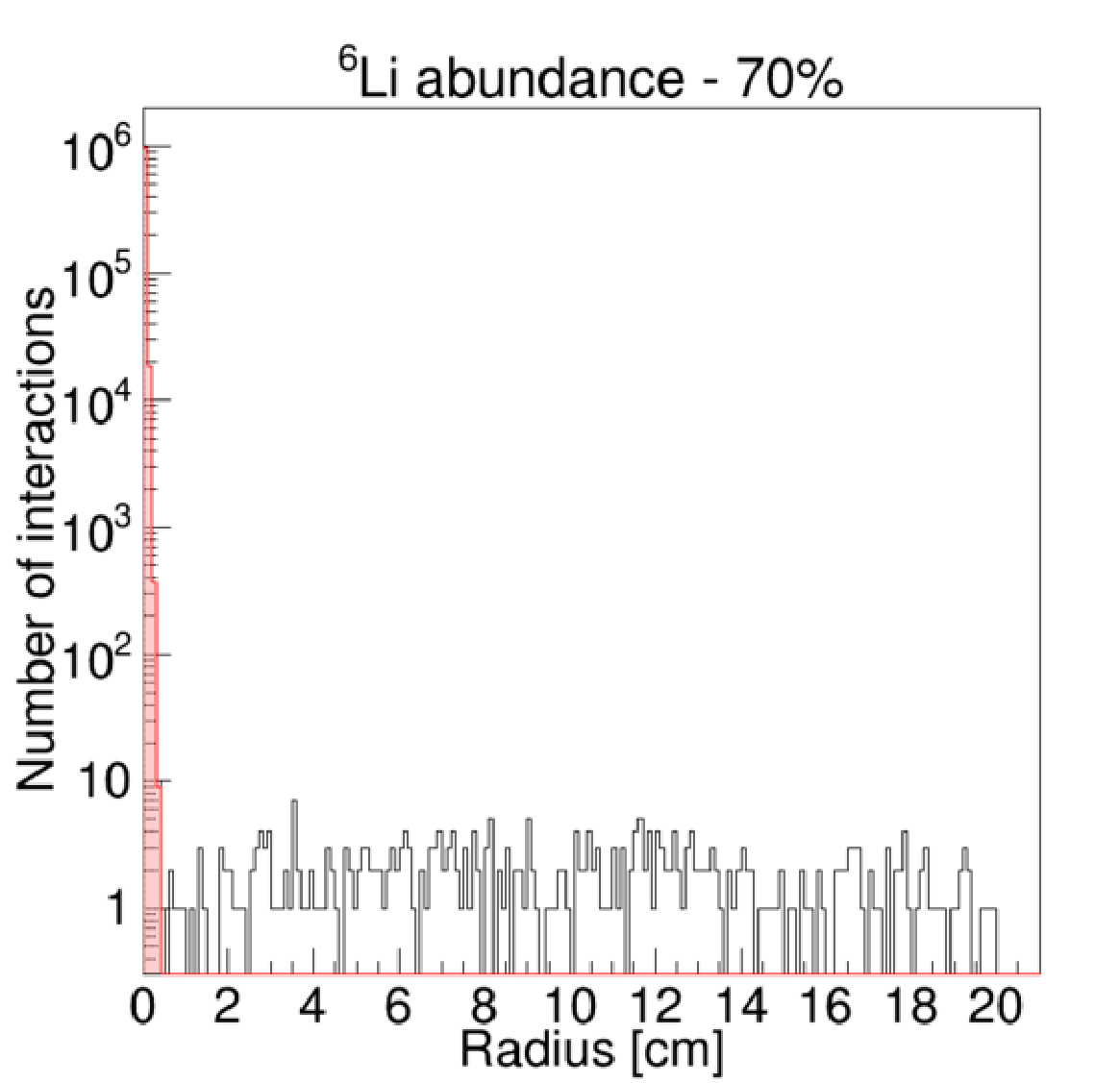} 
	\end{minipage}
		\hspace{0.5cm} 
	\begin{minipage}[t]{0.45\linewidth}
		\includegraphics[width= 0.9\textwidth]{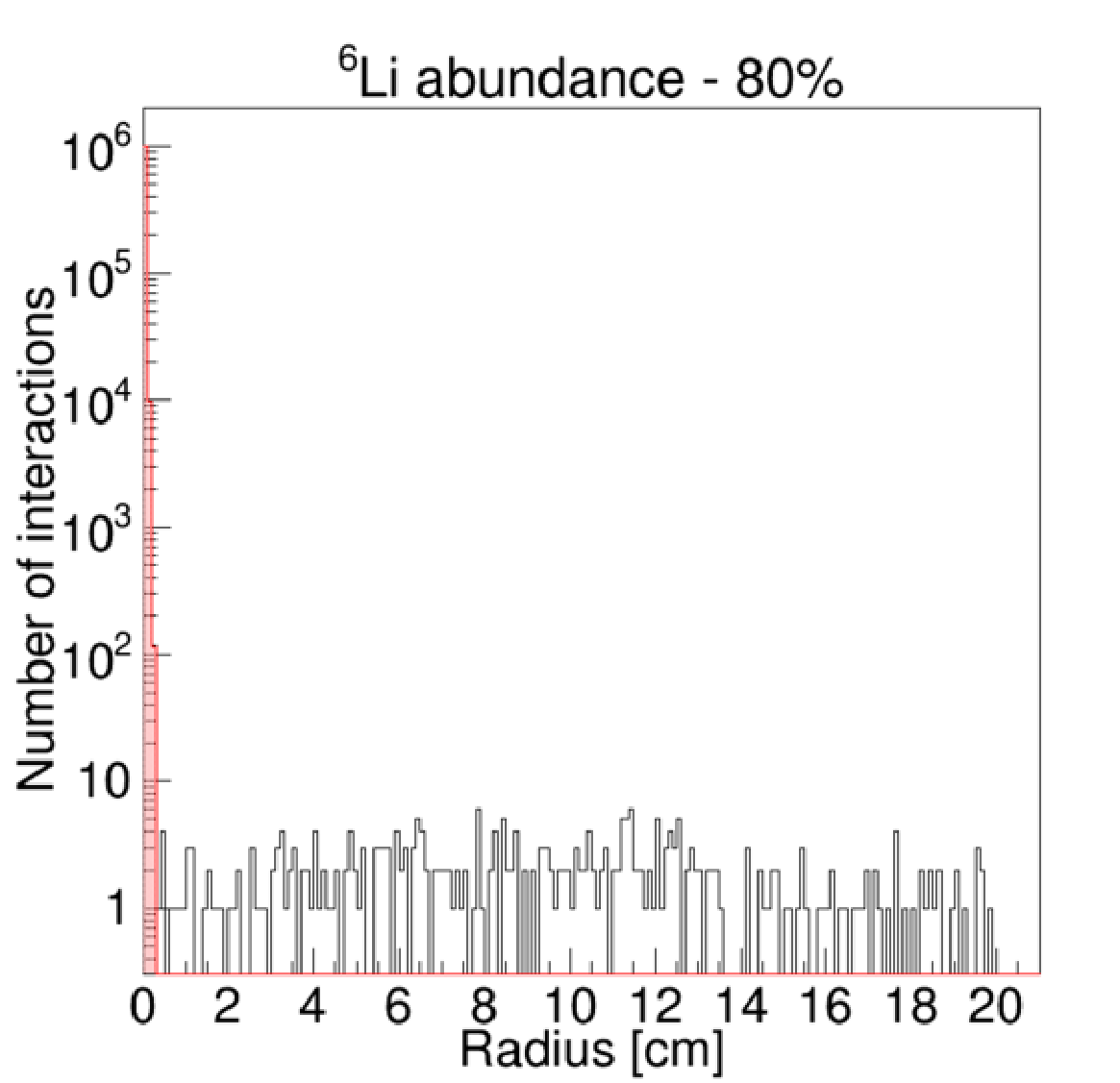} 	 
	\end{minipage}
 \end{Figure}

\begin{Figure}
\centering
	\begin{minipage}[t]{0.45\linewidth} 
		\centering
		 \includegraphics[width= 0.9\textwidth]{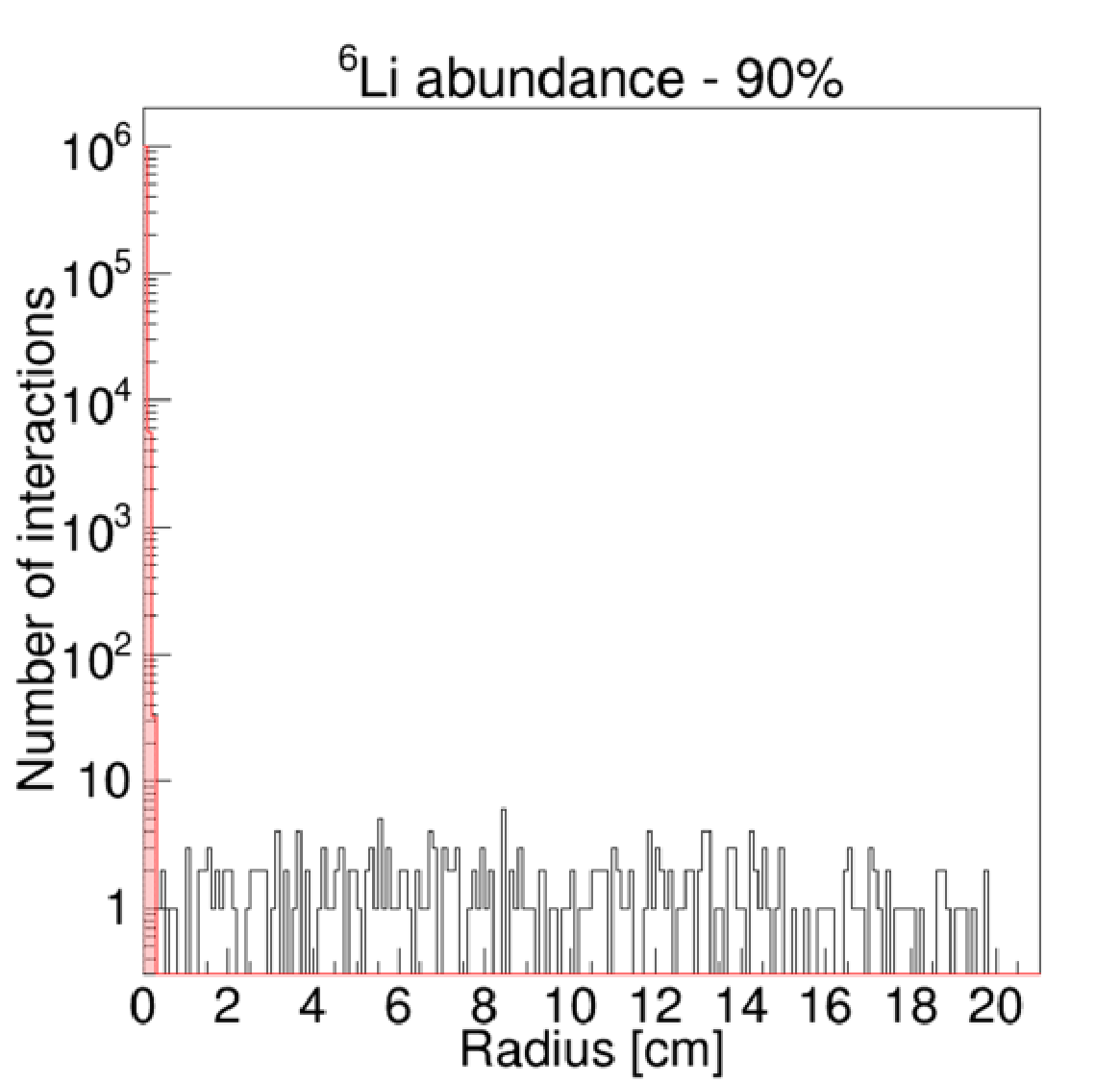} 
	\end{minipage}
		\hspace{0.5cm} 
	\begin{minipage}[t]{0.45\linewidth}
		\includegraphics[width= 0.9\textwidth]{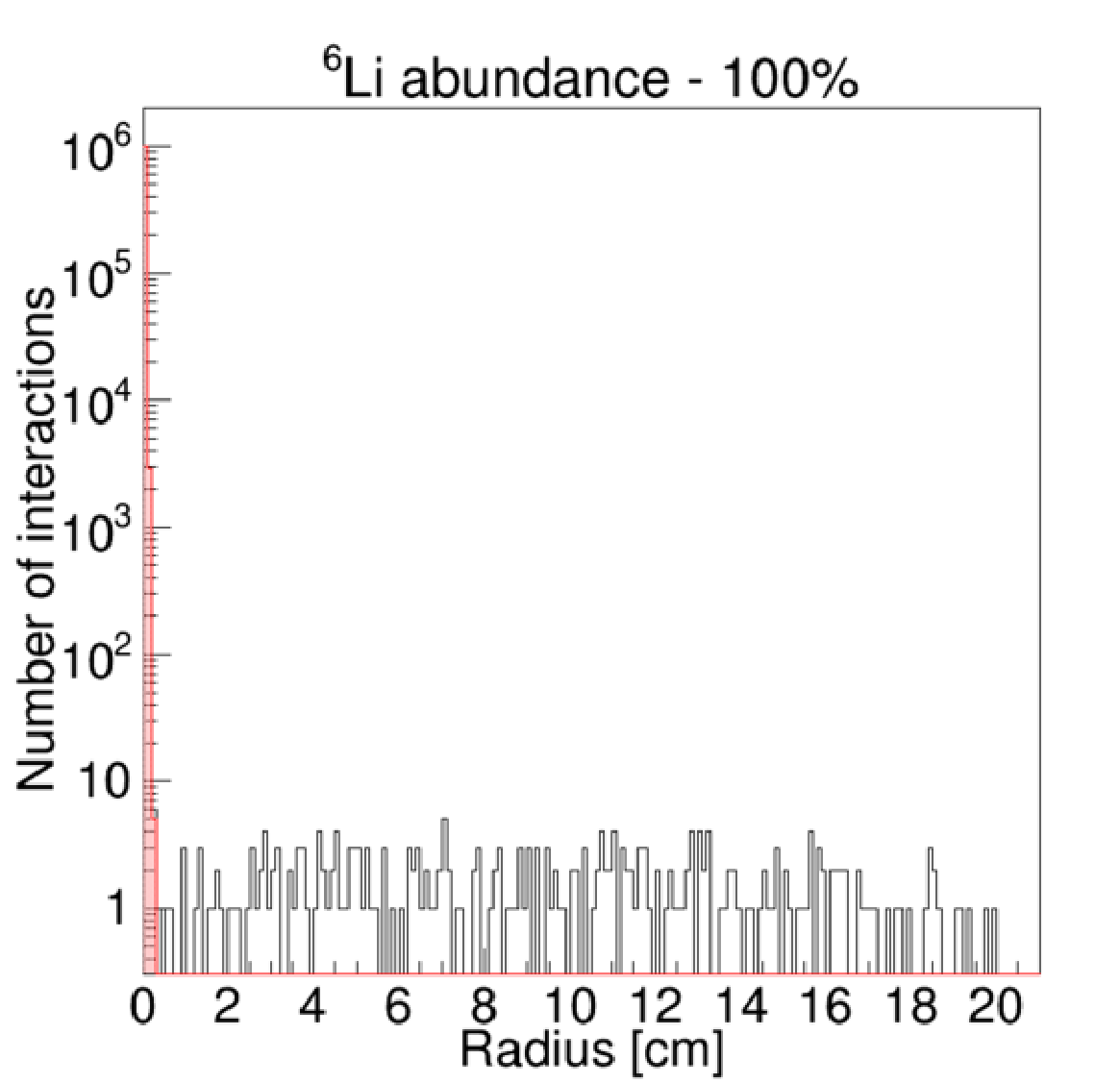} 	 
	\end{minipage}
	\captionof{figure}{Frequency of the reacion $^{6}Li + n \to  t + \alpha$ as a function of the fuel sphere radius, for Lithium-6 concentrations of 50\%, 60\%, 70\%, 80\%, 90\% e 100\%.}
\label{Spettri2}
 \end{Figure}

\begin{Figure}
	 \centering
	 \vspace{1.5cm}
 	\includegraphics[width=0.9\linewidth]{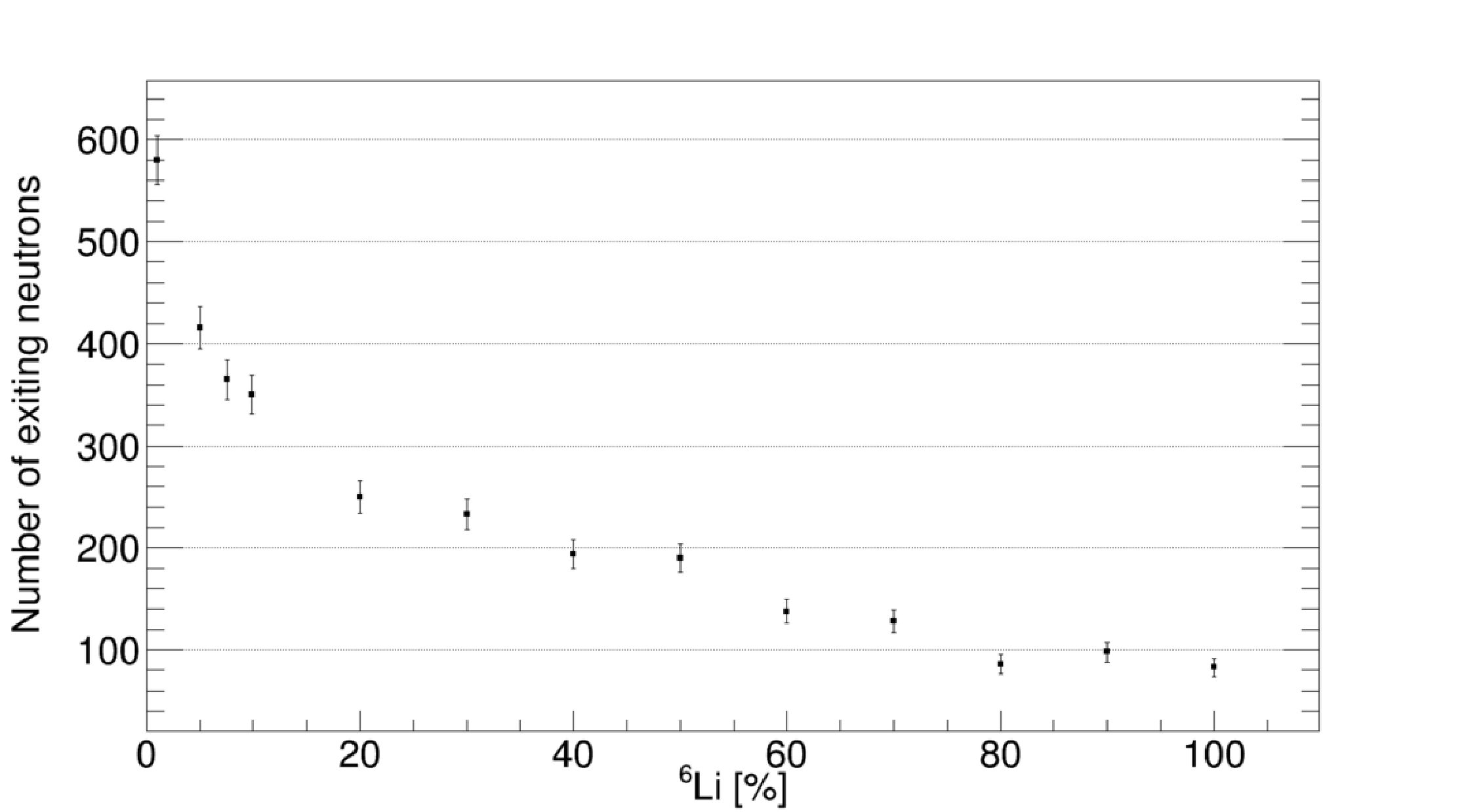}
 	\captionof{figure}{Number of neutrons exiting the fuel sphere as a function of Lithium-6 concentration. Number of primaries generated in all cases: $10^{6}$.}
 	\label{exitneu_number}
 \end{Figure}
 
\vspace{1.5cm}

\begin{Figure}
	 \centering
 	\includegraphics[width=0.9\linewidth]{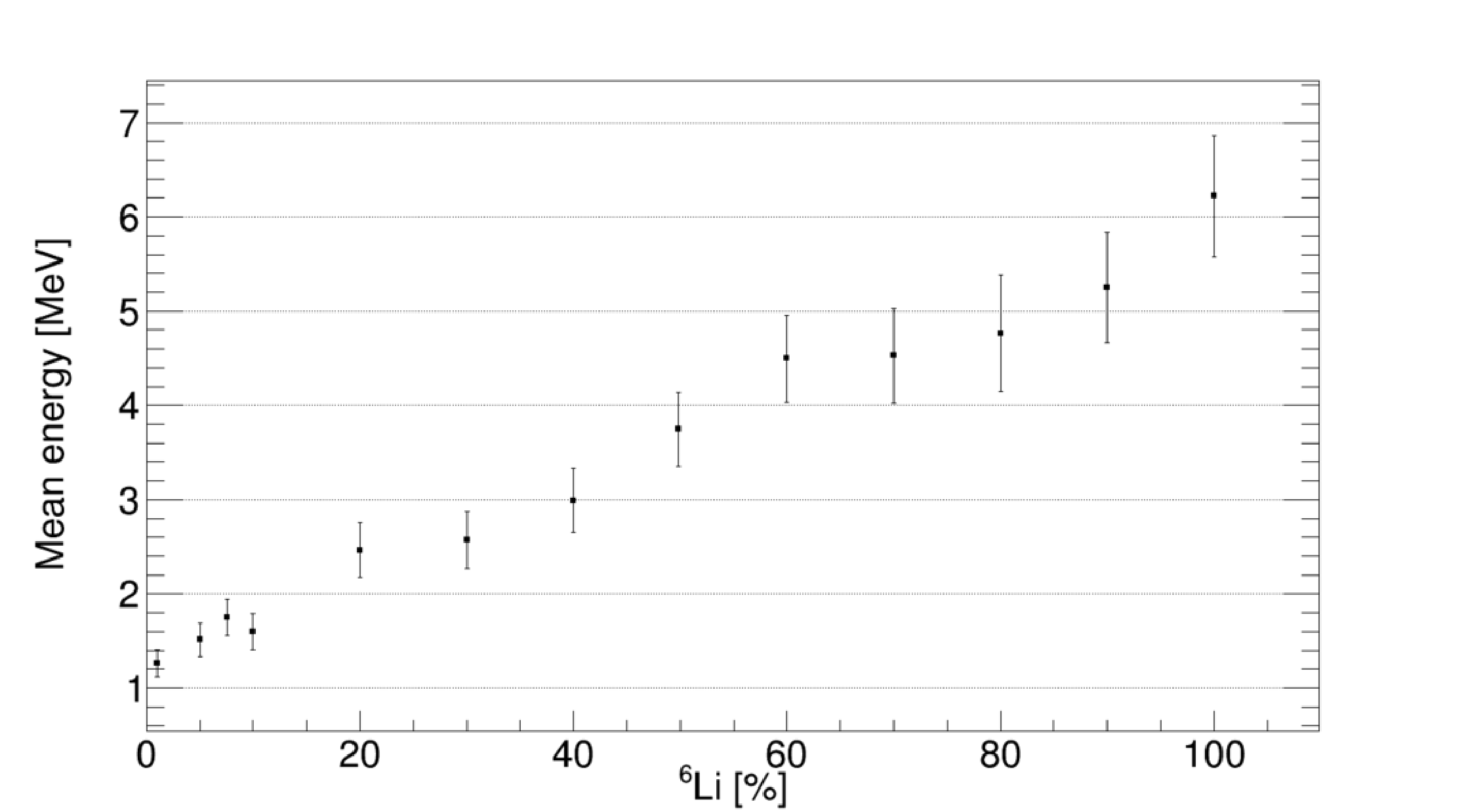}
 	\captionof{figure}{Mean energy of the neutrons exiting the fuel sphere as a function of Lithium-6 concentration.}
 	\label{exitneu_meanenergy}
 \end{Figure}

\newpage

 \begin{multicols}{2}

%

\subsection{Tritium Interaction in the Fuel Cell}
The interactions that Tritium can undergo within the fuel cell have also been analyzed. In most cases, it decays after a few elastic scatterings, while in a smaller number of cases, it interacts with Deuterium, Lithium-6, and Lithium-7. The frequencies of these interactions as a function of Lithium-6 concentration are illustrated in Figure~\ref{triton_process}. 

Figures~\ref{triton_etot_v2},~\ref{tine_alpha_nrg}, and~\ref{tine_neutron_nrg} respectively illustrate, for all interactions combined, as well as for the three most significant interactions:
\begin{equation}
		t + d \to  \alpha + n
\end{equation}
\begin{equation}
		t + ^{7}Li \to  2\alpha + 2n
\end{equation}
\begin{equation}
		t + ^{7}Li \to \textrm{other channels}
\end{equation}

\begin{itemize}
	\item The total energy available to the reaction products (Figure~\ref{triton_etot_v2});
	\item The energy released to $\alpha$ particles (Figure~\ref{tine_alpha_nrg});
	\item The energy released to neutrons (Figure~\ref{tine_neutron_nrg}).
\end{itemize}

By analyzing these graphs, one can deduce that Tritium interactions with Deuterium are independent of the Lithium-6 concentration (see the blue line in Figure~\ref{triton_process}, keeping in mind that, given the statistics of these events, each point has an error, not represented in the graph, between 5\% and 10\%). It has been verified that when interaction with Deuterium occurs, the energy available to the reaction products is approximately $19.5~MeV$ (see the blue distribution in Figure~\ref{triton_etot_v2}), and this energy is shared between the alpha particle and the neutron (see the blue distributions in Figures~\ref{tine_alpha_nrg} and~\ref{tine_neutron_nrg}, respectively).

Another observation from these graphs is that the primary Tritium reaction, occurring in about 55\%-60\% of cases for a natural concentration of Lithium-6, is with Lithium-7, producing 2 alpha particles and 2 neutrons (see the red line in Figure~\ref{triton_process}). In this case, the energy available to the reaction products is approximately $10.5~MeV$ (see the red distribution in Figure~\ref{triton_etot_v2}), and thus the produced alpha particles and neutrons have lower energy compared to the previous case (see the red distributions in Figures~\ref{tine_alpha_nrg} and~\ref{tine_neutron_nrg}, respectively).

Approximately an additional 20\% of Tritium interactions occur with Lithium-7 through different channels. In detail, all the observed reactions with the applied statistics ($10^{6}$ primary particles generated) are listed in Table~\ref{tabella_freq_proc_trizio}, along with their percentage frequency as a function of Lithium-6 concentration.

  \end{multicols}
  
  \vspace{2.5cm}
  
\begin{table}[h!]
\begin{center}
\resizebox{16cm}{!}{ 
\begin{tabular}{c|c|c|c|c|c|c|c|c|c|c|c|c|c|}
\cline{2-14}
 &  \multicolumn{13}{|c|}{$^6Li$ concentration [\%]} \\  
 \hline
\multicolumn{1}{ |c| }{Interaction} &    1  &  5  &  7.59  &  10  &  20  &  30  &  40  &  50  &  60  &  70  &  80  &  90  &  100  \\    
\hline
\multicolumn{1}{ |c| }{$ \textcolor{blue}{t + d \rightarrow \alpha + n}$}  &  \textcolor{blue}{18.3}  &  \textcolor{blue}{19.3}  &  \textcolor{blue}{19.3}  &  \textcolor{blue}{17.2}  &  \textcolor{blue}{18.2}  &  \textcolor{blue}{16.6}  &  \textcolor{blue}{17.7}  &  \textcolor{blue}{19.2}  &  \textcolor{blue}{17.9}  &  \textcolor{blue}{16.5}  &  \textcolor{blue}{16.3}  &  \textcolor{blue}{15.8}  &  \textcolor{blue}{18.7}  \\ 
\multicolumn{1}{ |c| }{$ \textcolor{red}{t + ^{7}Li \rightarrow 2\alpha + 2n}$} &   \textcolor{red}{65.3}  &   \textcolor{red}{56.4}  &   \textcolor{red}{53.6}  &   \textcolor{red}{59.1}  &   \textcolor{red}{46.4}  &   \textcolor{red}{46.9}  &   \textcolor{red}{40.0}  &   \textcolor{red}{30.7}  &   \textcolor{red}{26.2}  &   \textcolor{red}{19.8}  &   \textcolor{red}{12.1}  &   \textcolor{red}{7.8}  &   \textcolor{red}{0.0}  \\ 
\multicolumn{1}{ |c| }{$t + ^{7}Li \rightarrow n + ^{9}Be$}  &  6.6  &  5.8  &  7.0  &  5.8  &  6.7  &  5.9  &  4.7  &  5.0  &  1.0  &  2.3  &  2.0  &  1.0  &  0.0  \\ 
\multicolumn{1}{ |c| }{$t + ^{7}Li \rightarrow \alpha + ^{6}He$}  &  3.0  &  5.3  &  4.9  &  4.3  &  3.5  &  2.6  &  2.2  &  1.7  &  2.4  &  1.6  &  1.3  &  0.5  &  0.0  \\ 
\multicolumn{1}{ |c| }{$t + ^{6}Li \rightarrow d + ^{7}Li$}  &  0.2  &  1.4  &  1.0  &  1.5  &  2.7  &  6.2  &  6.6  &  7.0  &  9.0  &  14.0  &  14.6  &  17.1  &  18.7  \\ 
\multicolumn{1}{ |c| }{$t + ^{6}Li \rightarrow d + \gamma + ^{7}Li$} &  0.0  &  0.5  &  0.8  &  0.8  &  2.7  &  1.9  &  2.2  &  5.0  &  5.5  &  5.8  &  5.5  &  9.8  &  7.9  \\ 
\multicolumn{1}{ |c| }{$t + ^{6}Li \rightarrow t + ^{6}Li$}  &  0.0  &  0.0  &  0.0  &  1.3  &  2.1  &  2.8  &  2.5  &  3.6  &  4.5  &  4.2  &  5.0  &  2.8  &  6.3  \\ 
\multicolumn{1}{ |c| }{$t + ^{6}Li \rightarrow p + ^{8}Li$}  &  0.0  &  0.0  &  1.0  &  0.3  &  1.6  &  1.7  &  2.9  &  3.6  &  3.6  &  5.3  &  6.3  &  5.2  &  7.5  \\ 
\multicolumn{1}{ |c| }{$t + ^{7}Li \rightarrow t + ^{7}Li$}  &  5.4  &  7.0  &  8.0  &  4.8  &  5.1  &  3.8  &  3.4  &  2.4  &  2.1  &  2.8  &  0.5  &  0.5  &  0.0  \\ 
\multicolumn{1}{ |c| }{$t + ^{6}Li \rightarrow 2\alpha + n$}  &  0.5  &  2.8  &  3.9  &  4.3  &  10.7  &  10.7  &  16.0  &  19.7  &  26.4  &  25.6  &  34.7  &  37.5  &  39.7  \\ 
\multicolumn{1}{ |c| }{$t + ^{7}Li \rightarrow t + \gamma + ^{7}Li$}  &  0.7  &  1.6  &  0.3  &  0.8  &  0.3  &  0.7  &  1.0  &  1.0  &  0.7  &  0.2  &  0.0  &  0.3  &  0.0  \\ 
\multicolumn{1}{ |c| }{$t + ^{6}Li \rightarrow p + \gamma + ^{8}Li$}  &  0.0  &  0.0  &  0.3  &  0.0  &  0.0  &  0.2  &  0.7  &  1.2  &  0.7  &  1.9  &  1.8  &  1.8  &  1.2  \\

\hline
\end{tabular}        

} 
\end{center}
  \caption{The table shows the percentage frequencies of all Tritium interactions observed with the given statistics ($10^{6}$ primaries generated) as a function of the Lithium-6 concentration in Lithium Deuteride.}
   \label{tabella_freq_proc_trizio}
\end{table}

 \begin{Figure}
	 \centering
	 \vspace{1.5cm}
 	\includegraphics[width=0.9\linewidth]{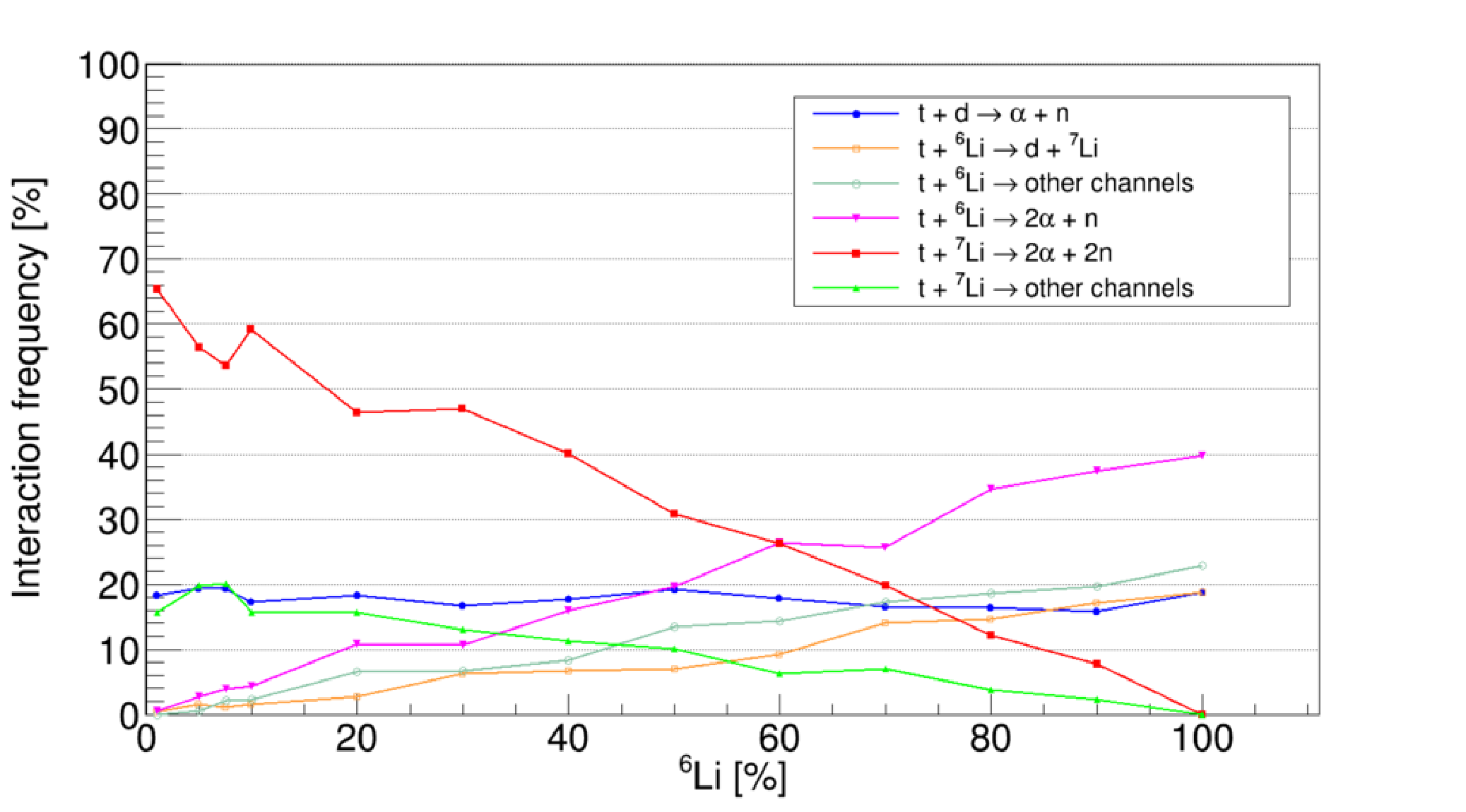}
 	\captionof{figure}{Frequency of various Tritium interactions in Lithium Deuteride as a function of Lithium-6 concentration.}
 	\label{triton_process}
 \end{Figure}
 
 \vspace{1.5cm}
 
\begin{Figure}
 	\centering
 	\includegraphics[width=0.9\linewidth]{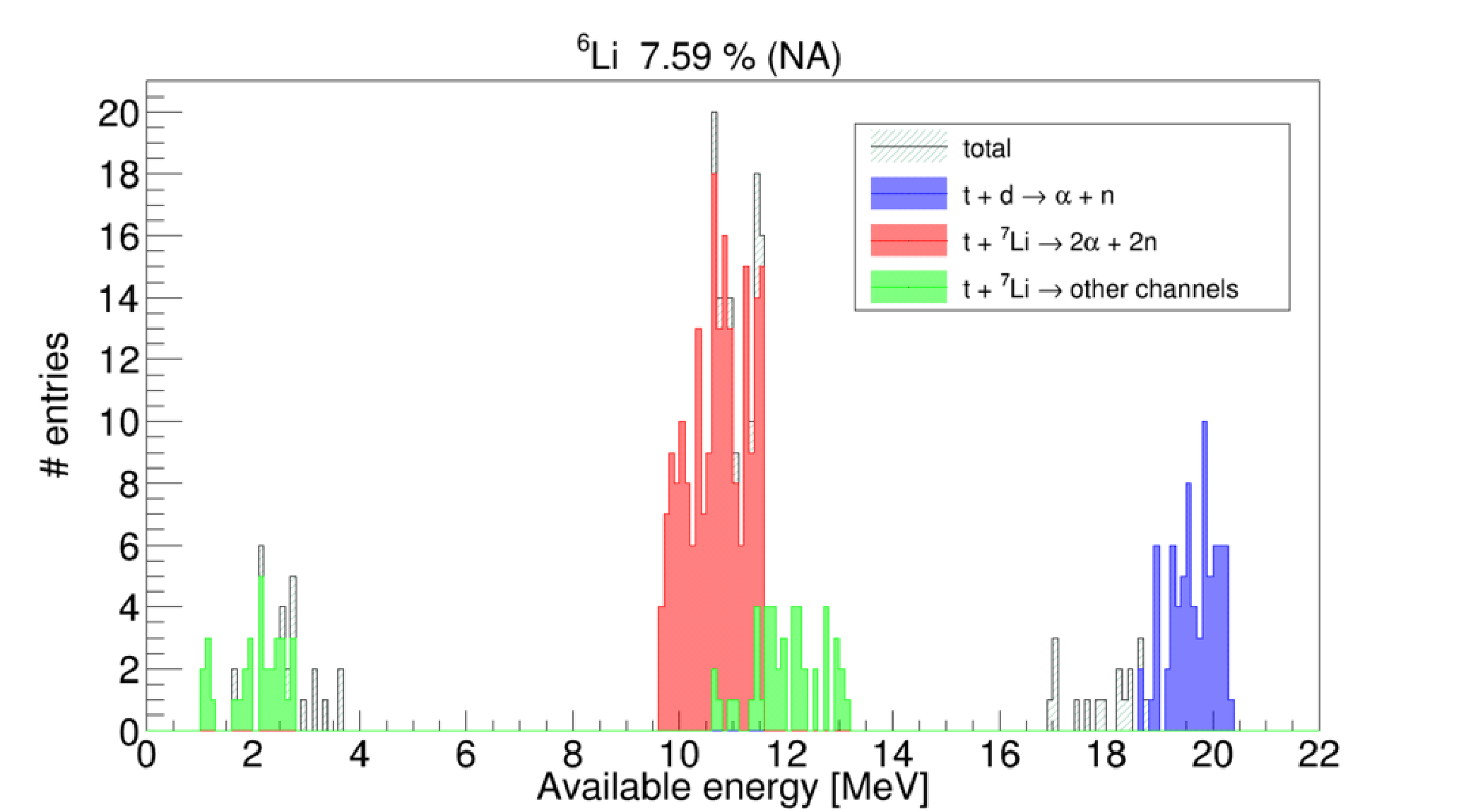}
 	\captionof{figure}{Spectra of the total energy available to the reaction products of the main Tritium reactions in Lithium Deuteride, with Lithium-6 at its natural concentration.}
 	\label{triton_etot_v2}
 \end{Figure}

\begin{Figure}
 	\centering
	\vspace{1.5cm}
 	\includegraphics[width=0.9\linewidth]{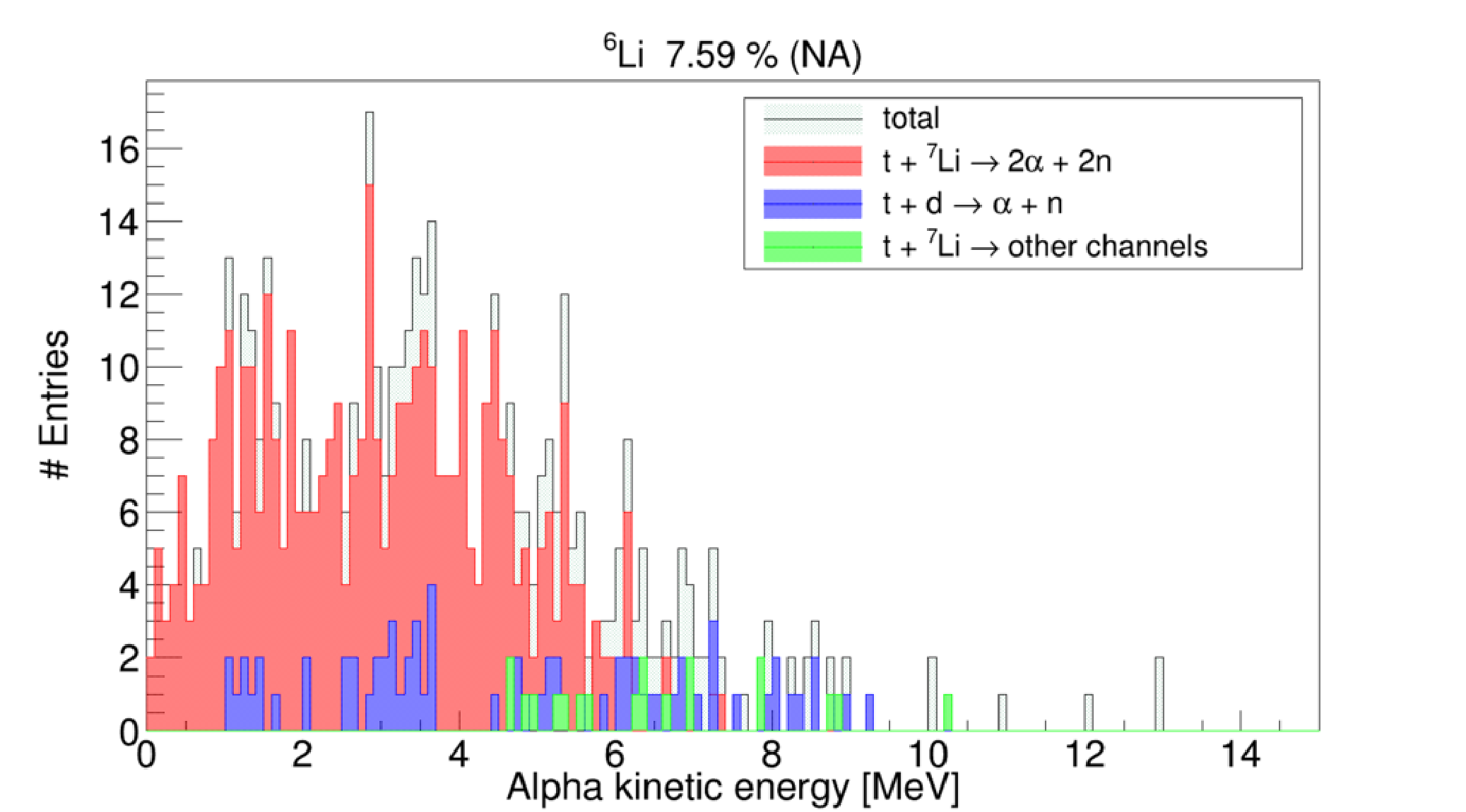}
 	\captionof{figure}{Spectra of the kinetic energy available to the alpha particles produced by the main Tritium reactions in Lithium Deuteride, with Lithium-6 at its natural concentration.}
 	\label{tine_alpha_nrg}
\end{Figure}

\vspace{1.5cm}

\begin{Figure}
 	\centering
 	\includegraphics[width=0.9\linewidth]{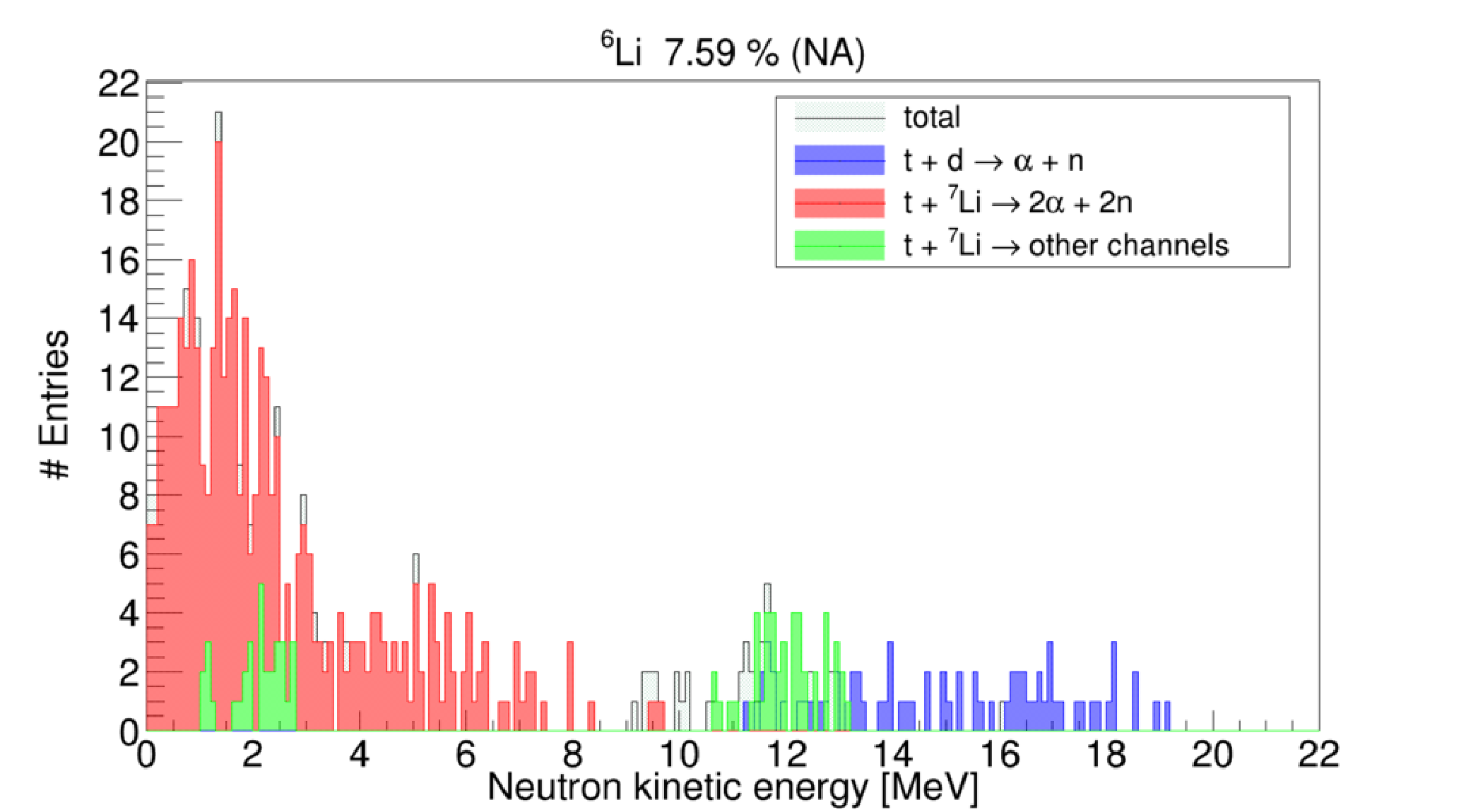}
 	\captionof{figure}{Spectra of the kinetic energy available to the neutrons produced by the main Tritium reactions in Lithium Deuteride, with Lithium-6 at its natural concentration.}
 	\label{tine_neutron_nrg}
\end{Figure}

\newpage

\begin{multicols}{2}
\section{Conclusions} \label{sec:CON}

From the results obtained in this study on the production and interactions of Tritium in Lithium Deuteride, the following considerations can be made:
\begin{enumerate}
 	\item In the first part, where the production of Tritium was analyzed, it was found that all primary neutrons generated are stopped within a few centimeters of the fuel sphere’s thickness, even with Lithium-6 at its natural concentration. Moreover, this stopping thickness rapidly decreases as the concentration of Lithium-6 increases. This suggests that it is not essential to achieve high enrichment levels, as even with just 10\%-20\% of Lithium-6, all primary thermal neutrons are stopped within less than 3~cm;
	\item In the second part, where the interactions of the produced Tritium were studied, it was confirmed that the process $t + d \to  \alpha + n$ is, as expected, independent of the Lithium-6 concentration. Therefore, this process does not require enrichment to be sustained. In addition to this process, another significant neutron-producing reaction was identified: $t + ^{7}Li  \to   2\alpha + 2n$, with an average available energy of approximately $10.5~MeV$. Although this energy is lower than the one of the other channel, it occurs with a higher probability.
\end{enumerate}

 \section*{Acknowledgments} \label{sec:RIN}
 We thank Futureon s.r.l. Centro Ricerche Energetiche, via Acqua Donzella 33, 00179 Rome, for supporting this work.

\end{multicols}
\end{document}